# Rattling-Induced Superconductiviy in the $\beta$-Pyrochlore Oxides $A$Os$_2$O$_6$

Yohei NAGAO, Jun-ichi YAMAURA, Hiroki OGUSU, Yoshihiko OKAMOTO, and Zenji HIROI[*]

*Institute for Solid State Physics, University of Tokyo, Kashiwa, Chiba 277-8581, Japan*

The superconducting properties of two $\beta$-pyrochlore oxides, CsOs$_2$O$_6$ and RbOs$_2$O$_6$, are studied by thermodynamic and transport measurements using high-quality single crystals. It is shown that the character of superconductivity changes systematically from weak coupling for CsOs$_2$O$_6$ to moderately strong coupling for RbOs$_2$O$_6$, and finally to extremely strong coupling with BCS-type superconductivity for KOs$_2$O$_6$, with increasing $T_c$. Strong-coupling correction analyses of the superconducting properties reveal that a low-energy rattling mode of the alkali metal ions is responsible for the mechanism of the superconductivity in each compound. The large enhancement of $T_c$ from Cs to K is attributed to the increase in the electron-rattler coupling with decreasing characteristic energy of the rattling and with increasing anharmonicity. The existence of weak anisotropy in the superconducting gap or in the electron-rattler interactions is found for the Cs and Rb compounds.

[*]E-mail: hiroi@issp.u-tokyo.ac.jp (footnote)

## 1. Introduction

Pyrochlore oxides with the general formula $A_2B_2O_7$ exhibit various intriguing phenomena such as magnetic frustration, a metal-insulator transition, colossal magnetoresistance, and ferroelectricity.[1] In 2001, superconductivity with $T_c$ = 1.0 K was found in cadmium rhenium oxide, Cd$_2$Re$_2$O$_7$, which was the first observation of superconductivity in the family of pyrochlore oxides.[2-4] Various experiments on Cd$_2$Re$_2$O$_7$ have revealed that the superconductivity is conventional and compatible with expectations based on the $s$-wave weak-coupling BCS theory.[5,6]

Later, alkali metal osmium oxides, $A$Os$_2$O$_6$ ($A$ = Cs, Rb, and K), were found to show superconductivity below $T_c$ = 3.3, 6.3, and 9.6 K for $A$ = Cs,[7] Rb,[8-10] and K,[11] respectively. This family of compounds is now called the $\beta$-pyrochlore oxides (Fig. 1), to distinguish them from the conventional pyrochlores with the general formula $A_2B_2O_7$, which are called the $\alpha$-pyrochlore oxides. Extensive study has been carried out on the three $\beta$-pyrochlore oxides in order to elucidate the nature of their superconductivity. For the superconducting gap symmetry, most experimental results indicate that it is of the $s$-wave BCS type,[12-20] suggesting conventional phonon-mediated superconductivity; there is no evidence of an alternative "glue" such as magnetic fluctuations, which are sometimes found in an exotic superconductor with a nodal gap. Interestingly, pronounced variations of various properties have been observed over the series.[21] It is obvious that a simple BCS picture cannot explain the systematic variation of $T_c$ from Cs to K, taking into account the similarity of the crystal and electronic structures over the series. There must be a special factor underlying the superconductivity of the $\beta$-pyrochlore oxides.

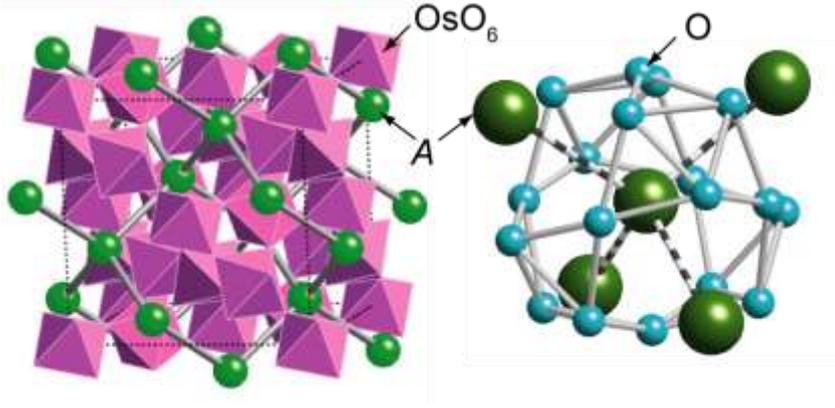

Fig. 1. (Color online) Crystal structure of the $\beta$-pyrochlore oxides $AOs_2O_6$. The $A$ atom is located in an oversized atomic cage made of $OsO_6$ octahedra and can move with a large excursion along the four [111] directions pointing to the neighboring $A$ atoms in adjacent cages.

It is a great advantage for studying this system that three compounds exist as well as another related superconductor, $Cd_2Re_2O_7$. We can investigate the chemical trends of various parameters over these compounds, which should lead to a deeper understanding of the underlying physics, after we have collected reliable experimental data for each compound. For $KOs_2O_6$, with the highest $T_c$, a high-quality single crystal was prepared and examined by various techniques; thus, the most important parameters have already been attained, revealing that superconductivity with extremely strong coupling is realized in $KOs_2O_6$.[12, 22] A large superconducting gap of the $s$-wave character with $2\Delta / k_BT_c$ = 4.5 - 5.1 was found by investigating the specific heat,[12, 22] photoemission spectra,[18] penetration depth,[16, 17] and thermal conductivity.[20] An NMR study using $^{17}O$ nuclei revealed the suppression of the Hebel-Slichter peak in the relaxation rate below $T_c$, which was ascribed to strong electron-phonon ($e$-ph) coupling.[23] A large upper critical field of above 30 T was estimated for $KOs_2O_6$.[22, 24, 25] Moreover, anomalous mixed states under magnetic fields were found and discussed in terms of vortex pinning.[26, 27] Furthermore, a first-order phase transition that occurs at $T_p$ = 7 - 8 K, which is below $T_c$, was found,[12, 17, 28, 29] the origin of which remains unclear.

In contrast, single crystals of $CsOs_2O_6$ and $RbOs_2O_6$ that are sufficiently large for physical characterizations have not yet been prepared, and most experiments have been performed on polycrystalline samples. Very recently, Legendre *et al*. reported the vortex phase diagram of $RbOs_2O_6$ using magnetization measurements on a single crystal.[30] It seems that their crystal has high quality but, surprisingly, exhibits a much lower $T_c$ of 5.5 K than the 6.28 K in the present work and the 6.3 - 6.4 K in previous studies on polycrystalline samples. Some thermodynamic parameters were extracted from specific heat measurements in the previous studies: the Sommerfeld coefficient $\gamma$ is approximately 40 mJ K$^{-2}$ mol$^{-1}$ for both compounds,[15] or 44 mJ K$^{-2}$ mol$^{-1}$ for Rb,[31] and the upper critical field $\mu_0 H_{c2}$ is 3.3 T and 5.5 - 10 T for Cs[21] and Rb,[8, 10, 21, 32, 33] respectively.  These results may contain some ambiguity, because the polycrystalline samples showed broad superconducting transitions, possibly originating from chemical inhomogeneity, crystallographic disorder, and impurity phases contained in the polycrystalline samples.  In particular, the large $H_{c2}$ values may not be intrinsic, because, in general, $H_{c2}$ can be enhanced by the strong scattering caused by disorder.  There is no doubt that higher-quality samples are required to obtain reliable data. Determining the superconducting properties of the two $\beta$-pyrochlore oxides, and thus deducing meaningful chemical trends over the series, would provide us with vital information on the mechanism of superconductivity in the $\beta$-pyrochlore oxides.

Recently, we succeeded in preparing high-quality single crystals of CsOs$_2$O$_6$ and RbOs$_2$O$_6$ by the chemical transport method, as shown in the photographs of Fig. 1. This allowed Terashima *et al.*[34] to observe the de Haas-van Alphen (dHvA) effect on one of the Cs single crystals and determine a Fermi surface (FS) that is consistent with the band structure calculations.  In this study, we carry out thermodynamic measurements using these single crystals in order to obtain basic parameters concerning the superconducting and normal states in a reliable manner.  Using the results, we discuss the chemical trends of various parameters over the whole series to explore the mechanism of superconductivity in the $\beta$-pyrochlore oxides.

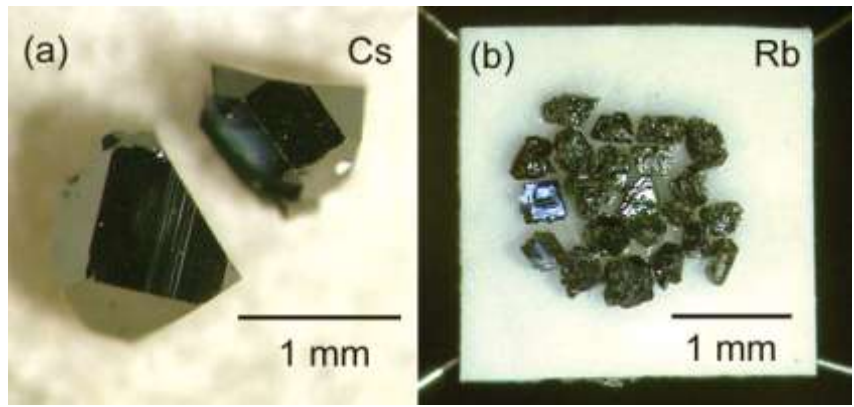

Fig. 2. (Color online) Photographs of single crystals of (a) CsOs$_2$O$_6$ and (b) RbOs$_2$O$_6$.  Rectangular (100) facets and triangular (111) facets can be seen on the large Cs crystals in (a).  Twenty-one tiny crystals of RbOs$_2$O$_6$ are placed on a square plate for specific heat measurements in (b).

An important structural feature to be focused on for the $\beta$-pyrochlore oxides is the rattling of the *A* atoms.  The $\beta$-pyrochlore oxides crystallize in a cubic structure with the space group *Fd*-3*m*, where a relatively small *A* cation lies at the 8*b* site in the center of an oversized atomic cage made of OsO$_6$ octahedra,[35, 36] as schematically depicted in Fig. 1. Although another space group, *F*-43*m*, was proposed for KOs$_2$O$_6$[37] and RbOs$_2$O$_6$,[38] recent reinvestigations by convergent-beam electron diffraction[39] and Raman spectroscopy[40, 41] have revealed that *Fd*-3*m* is the correct space group for all the compounds. Owing to the high symmetry of the site as well as the large size mismatch, the *A* atoms can "rattle" intensively inside the cage.[35, 42] The rattling has been recognized as a fascinating phenomenon for a class of cage compounds including filled skutterudites[43] and Ge/Si clathrates,[44] and has attracted many researchers because it may suppress thermal conductivity, leading to enhanced thermoelectric efficiency. Furthermore, the rattling itself is also intriguing from the viewpoint of lattice dynamics: it gives an almost localized mode even in a crystalline material and often exhibits unusual anharmonicity. Hence, it is considered that rattling is a novel type of low-lying excitation that potentially influences various properties of a crystal at low temperatures.

In the *β*-pyrochlore oxides, evidence of rattling has been obtained from structural refinements,[35, 36] specific heat measurements,[15, 22], NMR relaxation rates from the *A* nuclei,[23] and more recently, experiments using spectroscopic methods such as Raman[40, 41] and inelastic neutron scattering (INS).[45, 46] These experiments revealed large atomic displacements for the *A* cations and found low-energy excitations as well as their softening. For example, specific heat measurements reveal that the rattling mode is approximately described by the Einstein model with an Einstein temperature $\Theta_E$ of 70, 60, and 61 / 22 K for *A* = Cs, Rb, and K, respectively.[12, 15, 21, 22] Thus, the characteristic energy decreases with decreasing cation size or mass. This tendency illustrates the uniqueness of the rattling, because one would expect a higher energy for a lighter atom in the case of a conventional phonon. Moreover, it is remarkable that the specific heat shows an unusual $T^5$ dependence at low temperatures below 7 K for *A* = Cs and Rb, instead of the usual $T^3$ dependence expected from a normal Debye-type phonon.[15] On the other hand, the phase transition found only for the K compound can be related to the rattling degree of freedom.

We are interested in the possible interplay between the rattling of guest atoms and the conduction electrons moving around in the cage. There is a lot of experimental evidence of strong *e*-ph coupling especially in the case of $KOs_2O_6$. One piece of evidence is the concave downward temperature dependence of resistivity, suggesting an unusual scattering mechanism of the carriers.[12] Dahm and Ueda gave a theoretical explanation for this as well as the large enhancement of the K NMR relaxation rate, assuming the contribution of a local anharmonic phonon mode associated with the rattling of the K atom.[47, 48] Moreover, thermal conductivity,[20] penetration depth,[17] and photoemission spectroscopy (PES) measurements[18] revealed the dominant influence of the low-energy phonons on the transport properties of quasi-particles. Evidence of strong *e*-ph coupling is also found in the Raman spectroscopy data in the form of asymmetric peak broadening.[40, 49] It is of primary importance to clarify how the rattling affects the electronic properties and contributes to the mechanism of superconductivity in the *β*-pyrochlore oxides, which is the aim of this paper.

## 2. Experimental Procedure
### 2.1 Sample preparation

Single crystals of $CsOs_2O_6$ and $RbOs_2O_6$ were prepared by the following two-step reactions:

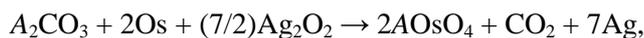
$$A_2\text{CO}_3 + 2\text{Os} + (7/2)\text{Ag}_2\text{O}_2 \rightarrow 2A\text{OsO}_4 + \text{CO}_2 + 7\text{Ag},$$

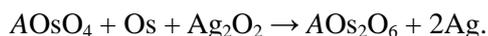
$$A\text{OsO}_4 + \text{Os} + \text{Ag}_2\text{O}_2 \rightarrow A\text{Os}_2\text{O}_6 + 2\text{Ag}.$$

The starting materials were $A_2\text{CO}_3$ ($A$ = Cs or Rb, Rare Metallic Co. Ltd., 99.9%) and Os (Alfa Aesar, 99.95%). First, $A\text{OsO}_4$ with a high oxidation state of $\text{Os}^{7+}$ was prepared by the reaction of $A_2\text{CO}_3$ and Os metal. The two powders were mixed in a molar ratio of 1:2, finely ground in an agate mortar, and pressed into a pellet in an argon-filled glove box. The pellet was placed in a silica tube, sealed under vacuum, and heated at 673 K for 24 h. An appropriate amount of $\text{Ag}_2\text{O}_2$, which decomposes into silver and oxygen at high temperatures above 673 K, was pelletized and placed in the same tube as an oxygen supplier. To avoid a direct reaction between the two pellets, they were separated by quartz wool. In the second step of the reaction, the obtained $A\text{OsO}_4$ and Os metal were mixed in an appropriate ratio, ground, and pressed into a pellet in the glove box. The pellet was again placed in an evacuated silica tube and heated at 793 - 833 K for 24 h, followed by slow cooling to room temperature.

After the reaction, many small crystals with an octahedral shape grew on the surface of the pellet in the case of $\text{CsOs}_2\text{O}_6$, while aggregates of tiny crystals were formed in the case of $\text{RbOs}_2\text{O}_6$. The maximum crystal size attained was approximately 1 and 0.1 mm, respectively. The crystal growth seems to proceed at temperatures near the melting point of $A\text{OsO}_4$ (833 K for $\text{CsOsO}_4$ and 793 K for $\text{RbOsO}_4$), although the growth mechanism is not clear.

In order to obtain larger and more reproducible crystals, we carried out crystal growth by a chemical transport method. A polycrystalline pellet of approximately 50 mg weight prepared by the second reaction was placed at one side of a sealed quartz tube of 150 mm length and 13 mm diameter. The tube was placed in a furnace with a temperature gradient, with the pellet at the higher-temperature side. Typical temperatures at either end were 920 and 890 K, so that the temperature gradient was ~0.2 K / mm. Crystals that were a few times larger than before grew on the wall of the tube at the low-temperature side, although it was still difficult to make Rb crystals as large as Cs crystals. It was noticed that the shape of the crystals changed from octahedral with (111) facets to cubic with (100) facets as they became larger (Fig. 2).

The metal composition of the thus-grown crystals was confirmed to be $A$ / Os = 0.50 ± 0.01 by energy-dispersive X-ray analysis using a scanning electron microscope. The as-grown crystals or polycrystalline samples may contain few vacancies at the $A$ site. However, we noticed that a certain number of vacancies were generated after the crystals were washed in water or an acid to remove the coexisting impurity phases followed by drying. Even if the number is small, it seems to broaden the superconducting transitions and in some cases lower $T_c$ considerably. In contrast to the case of $\text{KOs}_2\text{O}_6$, which is known to be hygroscopic,[50] the Cs and Rb compounds are more stable against moisture. The oxygen content could not be determined, and was assumed to be 7 per formula unit. It is likely from previous structural analyses by X-ray diffraction (XRD) and neutron diffraction (ND) that the nonstoichiometry of oxygen is negligibly small.[35, 36] The effects of disorder that can arise from chemical nonstoichiometry may not be serious for the present compounds.

The high degree of crystallinity was verified by the single-crystal XRD technique. Moreover, it was evidenced through resistivity measurements, as will be mentioned later, and recent dHvA experiments for a Cs crystal prepared by the same method.[34]

*2.2 Measurements of physical properties*

Resistivity measurements were carried out down to 0.5 K by the standard four-probe method in a Quantum Design physical property measurement system (PPMS). For $CsOs_2O_6$, a single crystal formed into a cuboid of $1.1 \times 0.3 \times 0.1$ mm$^3$ was used for the measurement, while, for $RbOs_2O_6$, an aggregate of tiny as-grown crystals was used, which might cause an ambiguity in evaluating the absolute value of resistivity, mainly owing to a dimensional error.

Heat capacity measurements were performed by the heat-relaxation method in the same PPMS. One crystal with a weight of 3.618 mg was used for Cs (Fig. 2(a)), and twenty-one small crystal with a total weight of 1.840 mg were examined for Rb. They were attached to an alumina platform by a small amount of Apiezon N grease, as shown in Fig. 2(b). The heat capacity of the addendum had been measured in a separate run without a sample and was subtracted from the data. Magnetic susceptibility was measured between 2 and 300 K in a Quantum Design magnetic property measurement system.

## 3. Results
*3.1 Superconducting properties*
*3.1.1 Resistivity*

Resistivity $\rho$ shows a sharp drop at a superconducting transition, as shown in Fig. 3. $T_c$, defined as an offset temperature, is 3.25 and 6.26 K for Cs and Rb, respectively. Over a wide range of $T$, each curve exhibits concave downward curvature above 70 K followed by saturation at approximately room temperature or $T^2$ behavior at low temperatures, as shown in Fig. 3(c). These $T$ dependences will be compared and discussed later for all the $\beta$-pyrochlore oxides, in relation to the Dahm-Ueda theory. Compared with the previous data obtained for polycrystalline samples,[7, 8] the overall $T$ dependence is similar for each compound, while the magnitude is much reduced, for example, it is one order smaller at room temperature for $CsOs_2O_6$. The value of $\rho$ at 300 K for Cs is 410 μΩ cm, which is close to the value expected for the Ioffe-Regel limit, which assumes a short mean free path (mfp) of carriers on the order of the interatomic distance of 0.36 nm due to strong scattering. In the case of $RbOs_2O_6$, $\rho$ is three times as large as that of $CsOs_2O_6$ at 300 K. However, this value may be an overestimate, mainly because of a dimensional error resulting from the irregular shape of the sample.

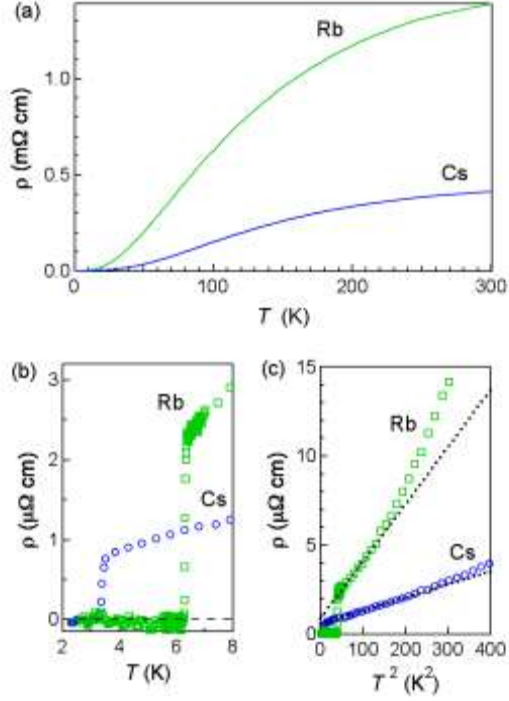

Fig. 3. (Color online) Resistivity $\rho$ measured on a reshaped crystal of $CsOs_2O_6$ and an aggregate of tiny as-grown crystals of $RbOs_2O_6$. (a) Temperature dependence of $\rho$ over a wide range, (b) superconducting transitions at low temperatures, and (c) $\rho$ versus $T^2$ plot. The broken lines in (c) are fits to the form $\rho = \rho_0 + AT^2$.

The extrapolation to $T = 0$ using a $T^2$ fit gives small residual resistivities $\rho_0 = 0.75$ μΩ cm for Cs and $\rho_0 = 1.1$ μΩ cm for Rb; thus, the residual resistivity ratios (RRRs) are very large, 550 and 1250, respectively. The RRR reported for a $KOs_2O_6$ single crystal was approximately 300.[12] These values are unusually large for metallic transition-metal oxides, demonstrating the high quality of our crystals. The mfp $l$ can be estimated using the relation

$$\rho_0 = \frac{\hbar(3\pi^2)^{1/3}}{e^2 \ell n^{2/3}}, \qquad (1)$$

where $n$ is the carrier density, which has not yet been determined experimentally. According to band structure calculations, $n$ is $2.8 \times 10^{21}$ cm$^{-3}$ (18% per Os),[21] almost independent of the system. Assuming this value, one obtains $l = 850$ nm for $CsOs_2O_6$, which is close to the corresponding values obtained from the dHvA experiments: 700 to 1000 nm depending on the band branches.[34] A similarly large value of $l \sim 580$ nm was estimated for $RbOs_2O_6$. Note that these values are comparable to that of a clean crystal of $Sr_2RuO_4$.[51]

*3.1.2 Magnetization*

A superconducting transition is also observed in the magnetization measurement of each compound. Each crystal exhibits a large shielding signal of over 100% and a Meissner signal corresponding to approximately 50%, as shown in Fig. 4. The shielding response of larger than 100% is due to the effect of demagnetization. The large Meissner response means inherently weak flux pinning in these three-dimensional superconductors. The onset temperature for a diamagnetic response is 3.36 and 6.24 K for Cs and Rb, respectively.

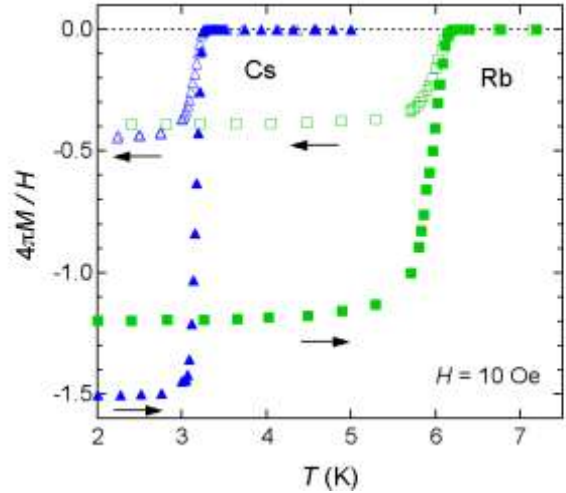

Fig. 4. (Color online) Magnetization measured at $H$ = 10 Oe on heating from 2 K after zero-field cooling and then cooling in the field. Large shielding and Meissner signals are observed for each compound.

*3.1.3 Specific heat*

Figure 5 shows the $T$ dependence of specific heat divided by $T$ at zero and high magnetic fields. In each compound, the superconducting transition at zero field is very sharp with a transition width of approximately 0.1 K. This is in strong contrast to the previous data on polycrystalline samples,[15] as compared in Fig. 6. The onset temperature is nearly equal for the single-crystalline and polycrystalline data, but the transition appears to collapse with a smaller jump at $T_c$ for the latter. This comparison illustrates how crucial the sample quality is for deducing reliable thermodynamic parameters from specific heat data. The sharp transition for the single crystal allowed us to determine the mean-field $T_c$ and the magnitude of the jump in specific heat at $T_c$ through an entropy-conserving construction assuming an ideal mean-field-type, second-order transition: $T_c$ = 3.25 K and $\Delta C / T_c$ = 61.5 mJ K$^{-2}$ mol$^{-1}$ for CsOs$_2$O$_6$, and $T_c$ = 6.28 K and $\Delta C / T_c$ = 81.8 mJ K$^{-2}$ mol$^{-1}$ for RbOs$_2$O$_6$. We examined three crystals for Cs and found scattering in $T_c$ within ±0.01 K and in $\Delta C / T_c$ within ±1 mJ K$^{-2}$ mol$^{-1}$. On the other hand, the transition observed for Rb is as sharp as that for Cs, in spite of the fact that the Rb sample consisted of 21 small crystals, which implies that the scattering in $T_c$ and $\Delta C / T_c$ for Rb must be smaller or similar to that for Cs. These facts indicate the absence of inherent defects or nonstoichiometry, which critically affects the superconductivity in the *β*-pyrochlore oxides. The values of $T_c$ obtained from the specific heat measurements are reasonably close to those obtained from resistivity and magnetic susceptibility measurements, and are taken as the most reliable values.

Superconductivity is suppressed completely under high magnetic fields, for example, 2 T for $CsOs_2O_6$ and 7 T for $RbOs_2O_6$, as shown in Fig. 5. Thus, we can safely determine the Sommerfeld coefficient $\gamma$ from the high-field data for each compound, in conjunction with the lattice contribution. Generally, the specific heat of a crystal ($C_p$) is the sum of an electronic contribution ($C_e$) and an $H$-independent lattice contribution ($C_l$). The former becomes $C_{en}$ for the normal state above $T_c$, which is taken as $\gamma T$, and $C_{es}$ for the superconducting state below $T_c$. $\gamma$ is assumed to be $T$-independent, although this may not be the case for compounds with strong $e$-ph coupling.[52] Two terms in the harmonic-lattice approximation are often required, which are sufficient for an adequate fit for $C_l$ at low temperatures; $C_l = \beta_3 T^3 + \beta_5 T^5$. The first term originates from a Debye-type acoustic phonon, and thus is dominant at low temperatures, while the second term expresses a deviation at high temperatures. Actually, this approximation is valid for $\alpha$-$Cd_2Re_2O_7$, where $\beta_3 = 0.222$ mJ K$^{-4}$ mol$^{-1}$ and $\beta_5 = 2.70 \times 10^{-6}$ mJ K$^{-6}$ mol$^{-1}$ have been obtained by a fit to the data below 10 K.[2] The Debye temperature $\Theta_D$ obtained from the value of $\beta_3$ is 458 K.

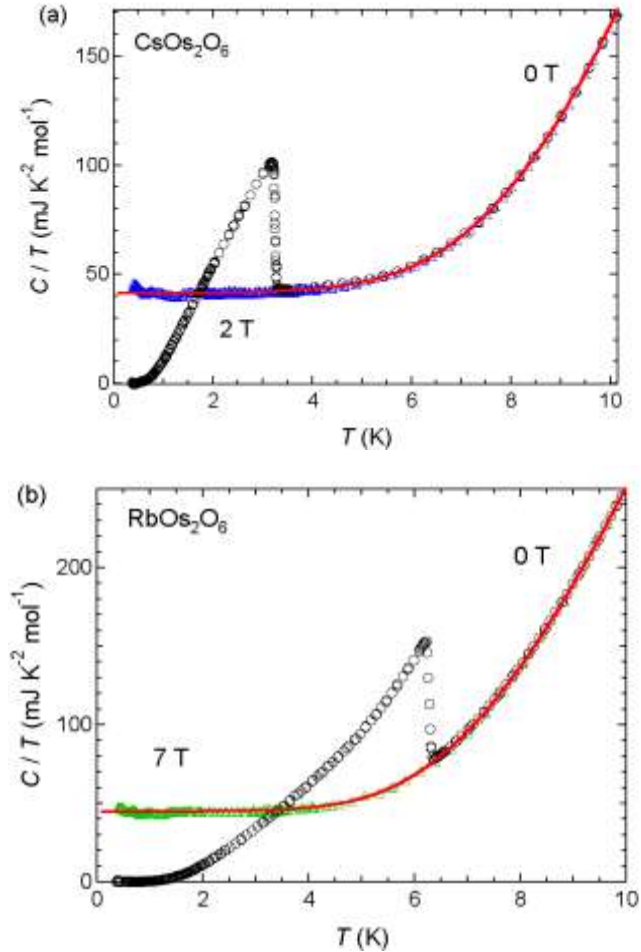

Fig. 5. (Color online) Specific heat divided by temperature, $C/T$, measured at $\mu_0 H = 0$ and 2 T for Cs (a) and at $\mu_0 H = 0$ and 7 T for Rb (b). The solid line in each panel is a fit to the form $C = \gamma T + \beta_5 T^5 + aC_E$.

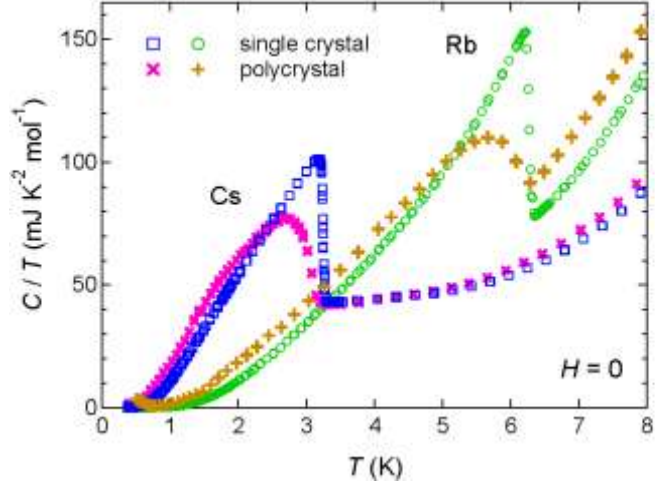

Fig. 6. (Color online) Comparison of superconducting transitions at $H = 0$ with changes in specific heat between the present single-crystalline and the previous polycrystalline samples.

In strong contrast, the $C/T$ at high fields shown in Fig. 5 for each compound becomes almost $T$-independent below ~4 K, suggesting that the first $T^3$ term is relatively small. In fact, fitting the data to the form $C = \gamma T + \beta_3 T^3 + \beta_5 T^5$ yielded a negative value for $\beta_3$. This is consistent with our previous results on polycrystalline samples, which indicate that the $T^5$ term is predominant even in this low-temperature range.[15] The origin of this unusual lattice contribution has not yet been understood and will be described in more detail elsewhere. Here we fit the data over a wide range of $T$ below 10 K to the form $C = \gamma T + \beta_5 T^5 + a C_E$, where $C_E$ is the contribution from an Einstein mode with Einstein temperature $\Theta_E$, and we obtain the following parameters: $\gamma = 41.38(2)$ mJ K$^{-2}$ mol$^{-1}$, $\beta_5 = 7.2(1) \times 10^{-3}$ mJ K$^{-5}$ mol$^{-1}$, $a = 0.344(8)$, and $\Theta_E = 66.1(4)$ K for CsOs$_2$O$_6$, and $\gamma = 44.73(6)$ mJ K$^{-2}$ mol$^{-1}$, $\beta_5 = 7.8(3) \times 10^{-3}$ mJ K$^{-5}$ mol$^{-1}$, $a = 0.47(1)$, and $\Theta_E = 57.1(2)$ K for RbOs$_2$O$_6$. Thus, the estimated values of $\gamma$ are greater than those reported previously of 40 mJ K$^{-2}$ mol$^{-1}$ for both Cs and Rb,[15] 34 mJ K$^{-5}$ mol$^{-1}$ for Rb,[10] and similar to the 44 mJ K$^{-2}$ mol$^{-1}$ reported for Rb.[10]

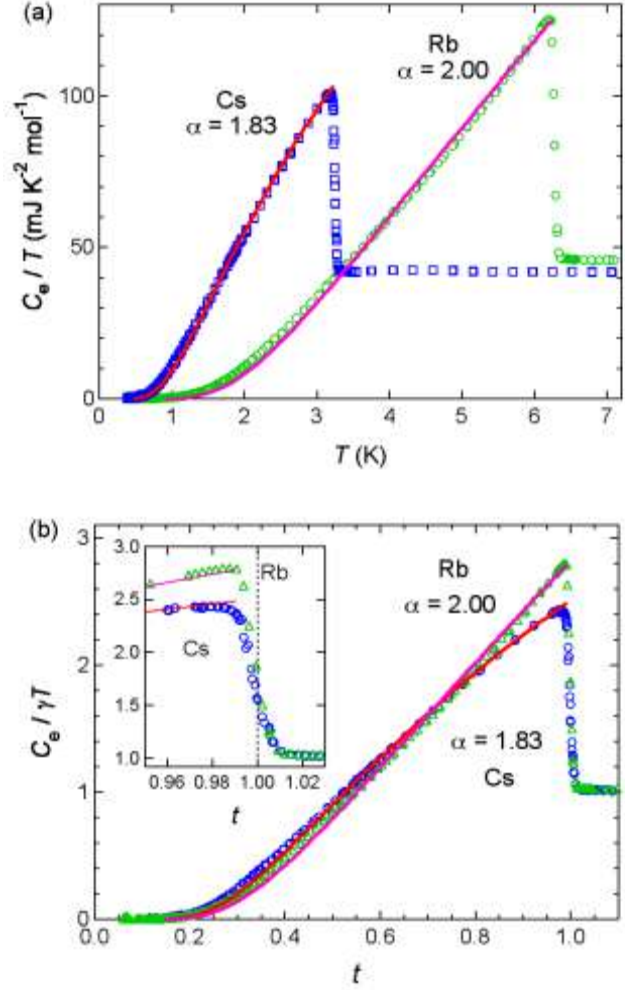

Fig. 7. (Color online) (a) Electronic specific heat divided by $T$, $C_e/T$, for $CsOs_2O_6$ and $RbOs_2O_6$. (b) Electronic specific heat divided by $\gamma T$ versus reduced temperature $t = T/T_c$. The inset enlarges the jump near $T_c$. The solid curve on each data set is a fit to the single-gap $\alpha$ model, which yields $\alpha$ = 1.83 and 2.00 for the Cs and Rb compounds, respectively.

Purely electronic specific heat was attained by subtracting the above lattice term and is plotted in Fig. 7. The jump at $T_c$, $\Delta C/\gamma T_c$, is 1.49 and 1.83 for Cs and Rb, respectively. The former is close to 1.43, as expected for a weak-coupling BCS superconductor, while the latter is significantly larger, although it is much smaller than 2.87 for $KOs_2O_6$, which lies in the extremely strong coupling regime.[12] As previously preformed for the K compound, we analyze the data of Cs and Rb on the basis of the $\alpha$ model, which has been developed to provide a semiempirical approximation to the thermodynamic properties of strong-coupling superconductors over a wide range of coupling strengths with a single adjustable parameter, $\alpha = \Delta_0/k_B T_c$.[53] As shown in Fig. 7, the $\alpha$ model fits the data of Cs with $\alpha$ = 1.83, which is slightly larger than the value for the weak-coupling BCS gap of 1.76. For Rb, the fitting gives a larger value of $\alpha$ = 2.00, which is close to the previously reported value of 1.94.[31] Although the fitting seems to be satisfactory for each compound, there is a discernible deviation between the experimental and calculated curves; the data points lie below the fitting curve near $T_c$ and above the curve at low temperatures below approximately half of $T_c$. Since the $\alpha$ model assumes the same $T$ dependence of the gap size as that predicted by BCS theory, such a deviation suggests possible anisotropy in the superconducting gap or the formation of additional gaps on different, minor Fermi surfaces, as in the cases of $MgB_2$[54] and $Nb_3Sn$.[55] For $KOs_2O_6$, on the other hand, a larger deviation from the fit to the $\alpha$ model was observed at low temperatures (see Fig. 21), although the second phase transition at $T_p$ made it difficult to analyze the data in detail.[12] It was suggested that this enhancement could be attributed to the coexistence of another smaller gap. Such a possibility has already been pointed out in a previous $\mu$SR experiment on $KOs_2O_6$.[19]

*3.1.4 Possible anisotropy in the superconducting gap*

The $T$ dependence of the electronic specific heat in the superconducting state, $C_{es}$, is shown in the Arrhenius plot of Fig. 8. $C_{es}$ cannot be approximated by a simple exponential form throughout the whole range of $T$ below $T_c$, and there appears to be an additional contribution at low temperatures. The magnitude of the superconducting gap at the low-temperature limit is estimated by fitting $C_{es}$ below $T_c / 3$ to an exponential form: $C_{es} = a\exp(-\Delta_0 / k_B T)$, as shown in Fig. 8. Fitting for Rb yields $\Delta_0 / k_B = 8.72(3)$ K, that is, $\alpha = 1.39$, which is 30% smaller than $\alpha = 2.00$ obtained from fitting by the $\alpha$ model over a wide range of $T$. Thus, as noted above, it is likely that there is anisotropy in the gap or that there are at least two gaps of different sizes. This is also the case for Cs: fitting the data below 1.2 K yields $\Delta_0 / k_B = 4.44(2)$ K, that is, $\alpha = 1.36$, which is 25% smaller than $\alpha = 1.83$ obtained from fitting by the $\alpha$ model. Therefore, a similar situation must occur in Cs and Rb, although it is less pronounced in Cs. Using the raw $C_p$ data instead of $C_e$ gave nearly the same results because of the negligibly small lattice contributions at low temperatures below $T_c / 3$. A similarly small gap of $\alpha = 1.6$ has been estimated for Rb from the $T$ dependence of the nuclear spin-lattice relaxation rate and was ascribed to anisotropy in the gap.[14] The present behavior of low-temperature specific heat resembles that reported for $NbSe_2$, where two gap values of $\alpha = 1.75$ and $0.16$ were clearly deduced for a high-quality crystal.[56]

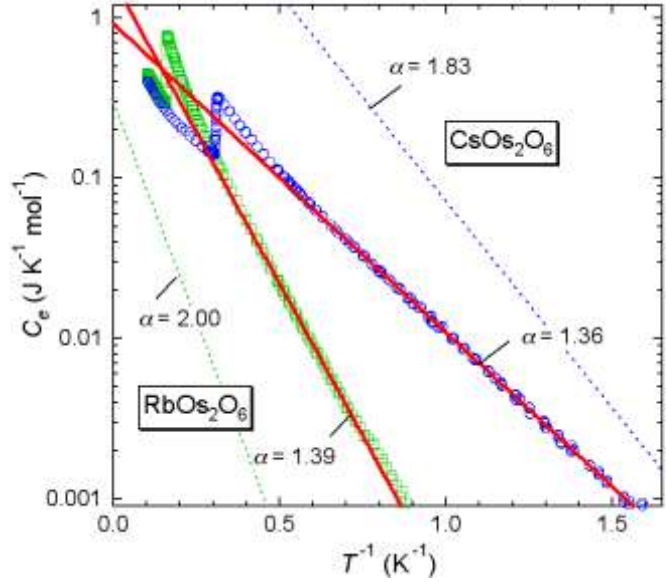

Fig. 8. (Color online) Electronic specific heat in the Arrhenius plot. The solid line on each data set represents a fit of the low-temperature data to an exponential form. The broken lines show slopes for $\alpha = 1.83$ (Cs) and 2.00 (Rb), which are predicted from the $\alpha$-model fitting of all the data shown in Fig. 7.

In order to obtain better fits to the specific heat data, a more elaborate analysis assuming two gaps is performed. The $\alpha$ model has been generalized to a multigap superconductor and successfully applied to analyses of $MgB_2$ and $Nb_3Sn$.[54, 55] In a two-band, two-gap model, the total specific heat can be considered as the sum of the contributions of each band calculated independently using the single-band model, assuming that interband transitions due to scattering by impurities or phonons can be neglected. Each band is characterized by a partial Sommerfeld constant $\gamma_i$ with $\gamma = \gamma_1 + \gamma_2$. Specific heat data are fitted with three parameters: the gap widths $\Delta_1$ and $\Delta_2$, and the relative weight $\gamma_1 / \gamma \equiv f$.[54] The results of fitting are improved significantly for Rb, as shown in Fig. 9(a), where we obtain $\alpha_1 = 1.6$, $\alpha_2 = 2.4$, and $f = 0.51$. The former is slightly larger than 1.39, obtained from the above-mentioned low-temperature limit, and the latter is larger than 2.0, obtained from the single-gap model. By carefully comparing the two fitting curves in Fig. 9(a) for the single and two-gap models, it is noticed that the steep decrease of $C_e$ immediately below $T_c$ requires a larger gap, while the gradual decrease below $t < 0.5$ can be reproduced by incorporating a smaller gap. Therefore, it is likely that two gaps with smaller and larger widths than $\alpha = 2.0$ open up with nearly equal weights. We cannot distinguish between the two possibilities of multigap and anisotropic single-gap superconductivity, both of which may give a similar correction to the single-gap $\alpha$ model. If there is anisotropy, one may interpret the above result to imply that the gap is approximately $1.6 < \alpha < 2.4$, as will be discussed later in further detail.

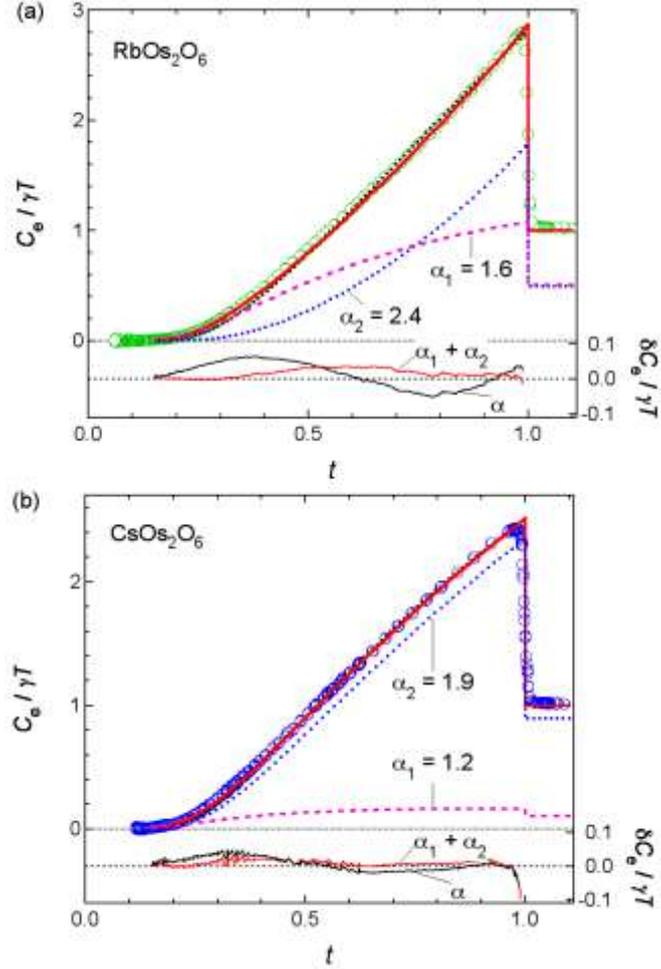

Fig. 9. (Color online) Fitting of $C_e/\gamma T$ with phenomenological models for Rb (a) and Cs (b). The dotted and solid lines near the data marks in each panel are fits to single-gap and two-gap $\alpha$ models with the two components shown by the lower broken lines. Deviations between experimental and calculated values are shown at the bottom for the two models. The parameters used for the two-gap fitting are (a) $\alpha_1 = 1.6$, $\alpha_2 = 2.4$, and $f = 0.51$ for Rb, and (b) $\alpha_1 = 1.2$, $\alpha_2 = 1.9$, and $f = 0.11$ for Cs.

In contrast, the same treatment for Cs results in less improvement in the fitting of the specific heat data, as shown in Fig. 9(b), as the single-gap model has already given a good fit. The estimated contribution from an additional small gap is rather small: $\alpha_1 = 1.2$, $\alpha_2 = 1.9$, and $f = 0.11$. Thus, it may be plausible for Cs that only a minor part of the FS possesses a relatively small gap size.

The $H$ dependence of the term linear in $T$ in the specific heat is shown in Fig. 10 for all the $\beta$-pyrochlore oxides. Generally, such a plot gives information about how a superconducting gap collapses by incorporating vortices and, thus, about the nature of the gap at $H = 0$. Typical behavior expected for an isotropic $s$-wave gap is $\gamma(H) \propto H$, whereas $\gamma(H)$ increases more rapidly with $H$ for the cases of a multiple gap, an anisotropic gap, and a gap with nodes. As shown in Fig. 10, all the data from the three $\beta$-pyrochlore oxides seem to fall on the same sublinear curve, suggesting the deviation from a simple isotropic gap. The initial rise at low fields is not steep, as for $MgB_2$[57] and $Y_2B_2C$,[58] but similar to that for NbSe$_2$.[58, 59] Thus, all the β-pyrochlore oxides must possess similar moderately anisotropic gaps. Note that such nonlinear behavior is observed only for a clean superconductor and becomes linear by intentionally adding impurities to Y$_2$B$_2$C and NbSe$_2$.[58] Thus, the nonlinear $\gamma(H)$ observed for the β-pyrochlore oxides implies superconductivity at the clean limit.

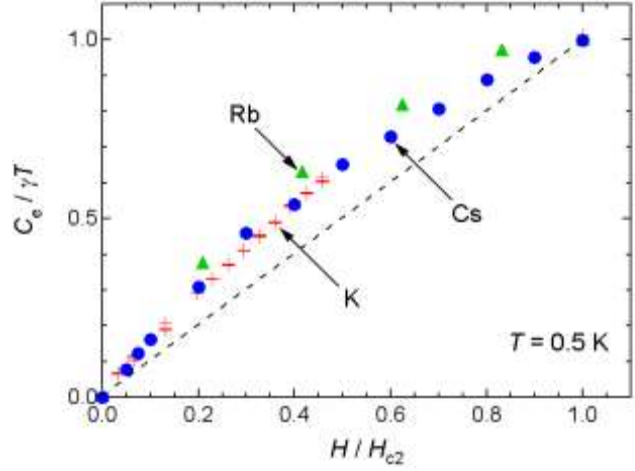

Fig. 10. (Color online) Magnetic-field dependence of the term linear in $T$ in the specific heat in the mixed state for the three β-pyrochlore oxides. The $C_e / T$ value at $T = 0.5$ K was taken to approximately represent the term linear in $T$ at $T = 0$. The broken line shows the proportional relation expected for an isotropic gap or a dirty superconductor.

*3.1.5 Thermodynamic critical field $H_c$*

The thermodynamic critical field $H_c$ is one of the most important parameters for characterizing the nature of the superconducting ground state, because it is related directly to the condensation energy as $\Delta F = H_c^2 / 8\pi$. It is deduced from specific heat data taken at $H = 0$ using the relation

$$H_c^2(T) = \frac{8\pi}{V_m} \int_T^{T_c} dT' \int_T^{T_c} dT'' \frac{C_{es}(T'') - C_{en}(T'')}{T''}, \quad (2)$$

where $V_m$ is the molar volume; $V_m = 78.772$ (77.962) cm$^3$ mol$^{-1}$ for Cs (Rb). The thus-determined $H_c(T)$ for each compound is shown in Fig. 11. The validity of this procedure is verified using the Rutgers' equation

$$\frac{1}{V_m} \frac{\Delta C}{T_c} = \frac{1}{4\pi} \left( \frac{dH_c}{dT} \right)^2_{T_c}. \quad (3)$$

The initial slope of $H_c$ at $T_c$ is -30.3 (-36.5) mT K$^{-1}$ for Cs (Rb) from the data in Fig. 11 and is compared with the value calculated by the Rutgers' equation using $\Delta C / T_c = 61.5$ (81.8) mJ K$^{-2}$ mol$^{-1}$; $-(dH_c / dT)_{T_c} = 31.3$ (36.3) mT K$^{-1}$. Thus, our result is in good agreement with those calculated by the Rutgers' equation for each compound. The magnitude of a superconducting gap can also be estimated using the initial slope obtained from the Toxen relation[60]

$$-\frac{2T_c}{H_c(0)} \frac{dH_c(T)}{dT} \bigg|_{T_c} = \frac{2\Delta(0)}{k_B T_c}, \quad (4)$$

which yields $2\Delta(0) / k_B T_c = 3.33(3.64)$ for Cs (Rb) and 4.00 for K. The equation seems to underestimate the gap size, particularly for K,[12] suggesting that it may not be applicable to the strong-coupling case.[61]

As shown in the inset of Fig. 11, $H_c$ follows the parabolic form $H_c = H_c(0)(1 - t^2)$, where $t = T / T_c$, at low temperatures. $H_c(0)$ is determined unambiguously from an extrapolation to $T = 0$, and $H_c(0) = 58.96(2)$ mT for Cs and $125.85(3)$ mT for Rb. $H_c(0)$ for Rb agrees well with a previous value given by Brühwiler *et al.*; $H_c(0) = 124.9$ mT.[31]

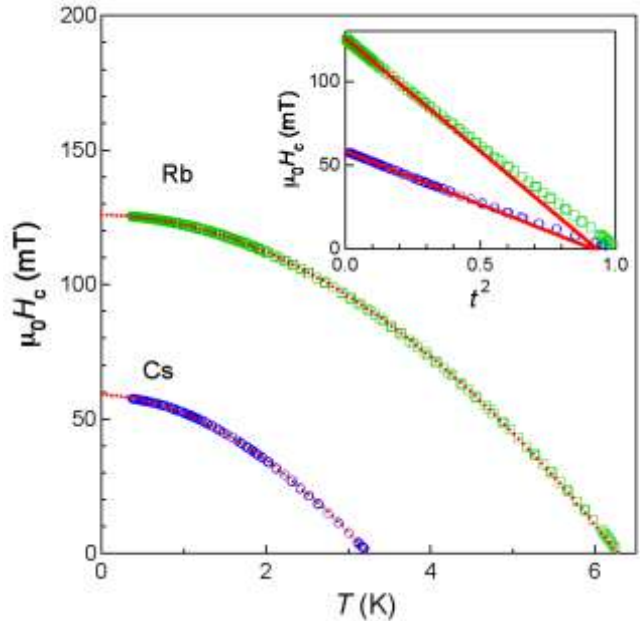

Fig. 11. (Color online) Thermodynamic critical field $H_c$. The inset shows $H_c$ vs $t^2$, where the solid line on each data set is a fit to the form $H_c = H_c(0)(1 - t^2)$ in the low-temperature limit. The broken curves in the main panel represent the same function with the values of $H_c(0)$ determined from the fits in the inset.

The deviation from the parabolic $T$-dependence of $H_c$ is often used to characterize the superconducting state by considering the deviation function $D(t) = H_c(t)/H_c(0) - (1 - t^2)$. It is always negative for a weak-coupling BCS superconductor such as Al and positive for a strong-coupling one such as Pb, as shown in Fig. 12. In the case of intermediate coupling, it is small and sometimes changes its sign from negative to positive with decreasing $t$, as observed in the cases of V$_3$Si and CeRu$_2$.[62] The $D(t)$ of CsOs$_2$O$_6$ shown in Fig. 12 is always negative and close to the BCS curve. In contrast, the behavior observed for RbOs$_2$O$_6$ is rather strange: it is positive at high temperatures but becomes negative on cooling. This unusual behavior may be related to the multigap or the anisotropic nature of the gap for Rb, as suggested from the analysis of the specific heat. For KOs$_2$O$_6$, which is a strong-coupling superconductor, $D(t)$ is positive and close to that of Pb, although the availability of data is limited above 8.5 K.[12]

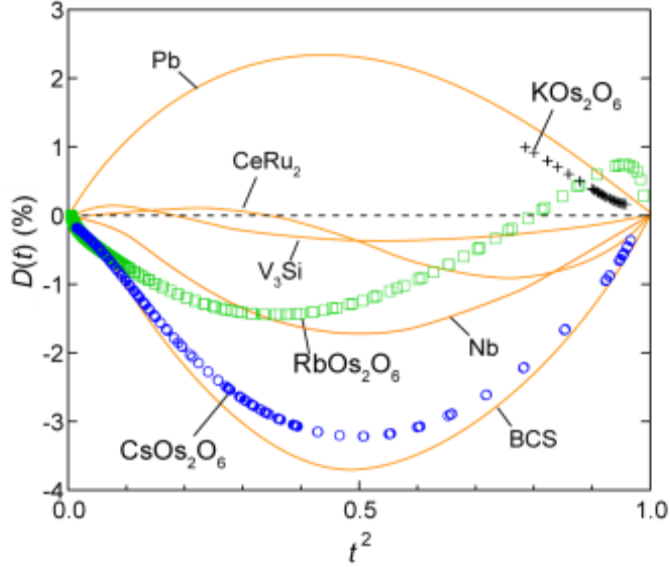

Fig. 12. (Color online) Thermodynamic critical field deviation function $D(t)$ as a function of the square of the reduced temperature $t$ for $CsOs_2O_6$ and $RbOs_2O_6$ in comparison with that for $KOs_2O_6$[12] and other typical superconductors, Pb, Nb, $V_3Si$, and $CeRu_2$.[62] The lowest curve is the BCS result.

Our $D(t)$ curve for Rb is considerably different from that reported by Brühwiler et al.,[31] which seems more conventional and has been reproduced within the framework of s-wave Eliashberg theory, which includes anisotropy and the effects of impurity scattering.[63] However, it is noticed that there is a similar tendency for their $D(t)$ curve to become positive near $T_c$ as in our curve. The experimental procedure to determine $D(t)$ is sensitive, particularly near $T_c$, to the sharpness of the transition in the specific heat data. We consider that our $D(t)$ data is more reliable, because our crystal exhibits a much sharper transition of $\Delta T_c \sim 0.1$ K than their polycrystalline sample, for which $\Delta T_c \sim 0.4$ K. Moreover, the effects of impurity scattering are expected to be critical in their sample, which may obscure the possible anisotropy in the gap, if it exists.

*3.1.6 Upper critical field $H_{c2}$*

The evolution of superconducting transitions with changes in the specific heat under various magnetic fields is shown in Fig. 13, where the jump at $T_c(H)$ systematically becomes smaller and shifts to lower temperatures with increasing $H$. The jump remains sharp under magnetic fields with less broadening compared with the cases of polycrystalline samples.[15] The $H$ dependence of $T_c$ or the $T$ dependence of the upper critical field $H_{c2}$ is plotted in Fig. 14, together with data from resistivity measurements using different crystals from the same batch. In each compound, $H_{c2}$ increases with concave upward curvature near $T_c$ and saturates at low temperatures. The initial slope at $T_c$ is $-d(\mu_0 H_{c2})/dT = 0.44$ and $0.88$ T K$^{-1}$ for Cs and Rb, respectively. The value for Rb is smaller than a previously reported value of 1.2 T K$^{-1}$.[31] Overall, the values of $H_{c2}$ are also considerably smaller than those previously reported.

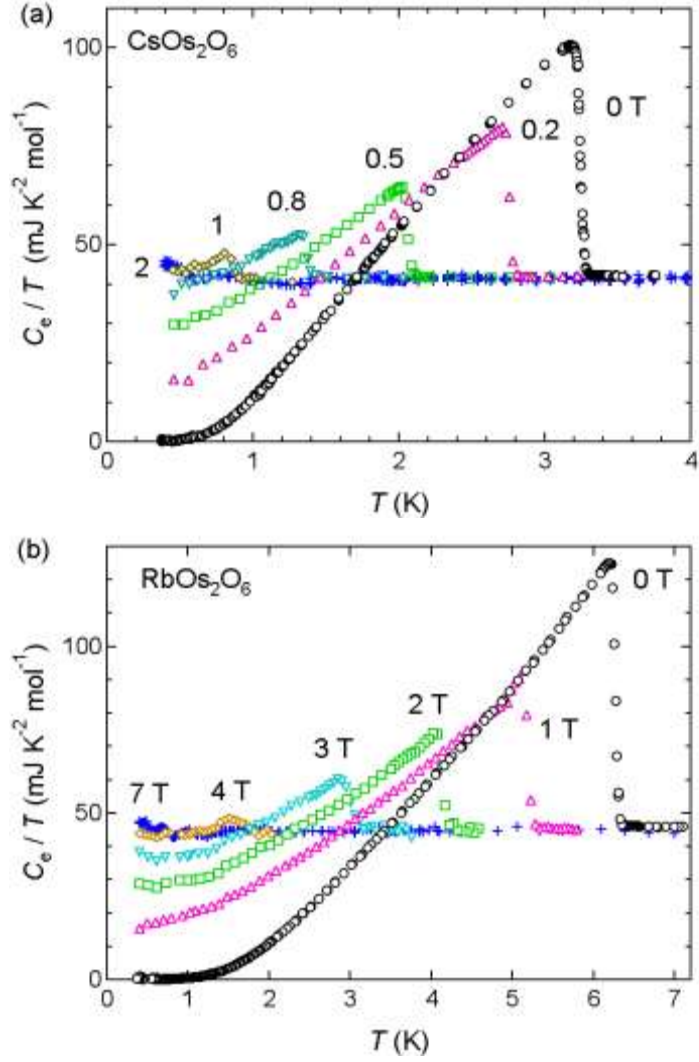

Fig. 13. (Color online) Superconducting transitions under magnetic fields with changes in specific heat for (a) Cs and (b) Rb. The magnitudes of the field are (a) 0, 0.2, 0.5, 0.8, 1, and 2 T and (b) 0, 1, 2, 3, 4, and 7 T from right to left.

Here it is worthwhile discussing the variation of the initial slope from Cs to K, as already discussed by Brühwiler et al.[33] The slope $-d(\mu_0 H_{c2})/dT|_{Tc}$ increases with values of 0.44, 0.88, and 3.61 T / K for Cs, Rb, and K, respectively, in a ratio of approximately 1:2:8. It is known that the slope is proportional to $(\tau v_F^2)^{-1}$, where $\tau$ is the electron scattering time and $v_F$ is the Fermi velocity. Since the bare $v_F$ obtained from band structure calculations is nearly the same over the series ($\sim 2.6 \times 10^5$ m/s),[33] the difference should originate from the renormalization of $v_F$ and $\tau$. Taking $v_F$ as $v_F / (1 + \lambda)$, where $1 + \lambda = \gamma_{exp} / \gamma_{band}$ is 3.76, 4.38, and 7.3 for Cs, Rb, and K, respectively, as listed in Table I, the ratio of $\tau$ becomes 1:0.68:0.46. This implies the enhancement of carrier scattering from Cs to K, suggesting stronger e-ph coupling with increasing $T_c$. If one takes $l_{Cs} = 850$ nm, estimated from $\rho_0$, we obtain $l_{Rb} = 580$ nm and $l_K = 390$ nm. The value for Rb is equal to that estimated from $\rho_0$, while $l_K = 390$ nm is much smaller than $l_K = 640$ nm estimated from $\rho_0$ in the case of K. This means that strong scattering still occurs for K even at $T_c$, as also suggested by the surviving concave downward resistivity[12] as well as the observations of short-lived quasi-particles by PES[18] and thermal conductivity measurements.[17]

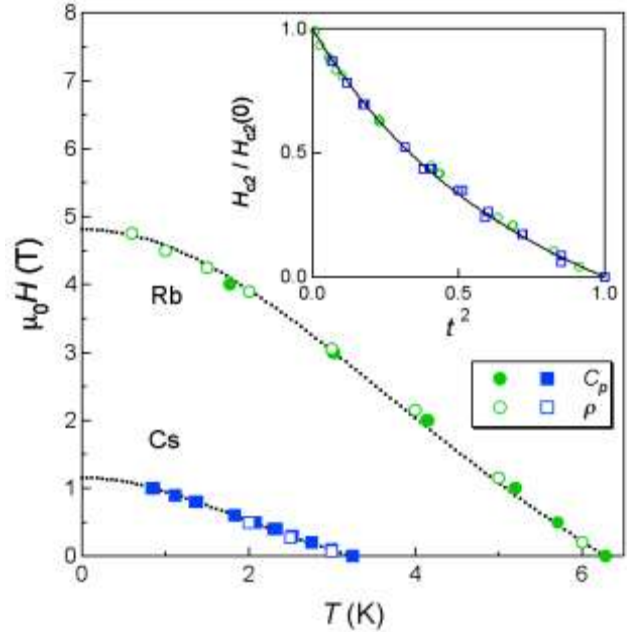

Fig. 14. (Color online) Upper critical field $H_{c2}$ determined by specific heat (closed symbols) and resistivity measurements (open symbols). The inset shows $H_{c2}$ normalized by $H_{c2}(0)$ vs $t^2$. The solid curve in the inset and the broken curves in the main panel represent fits to eq. (6).

The initial upward curvature of the $H_{c2}$ line is similar to that reported for pure Nb, which was ascribed to FS anisotropy.[64] It was also reported for Nb that the curvature disappeared to show a linear response with increasing impurity scattering.[64] Similarly, in the present compounds, the effects of FS anisotropy may exist and manifest themselves in the upward curvature, reflecting the good sample quality.

According to the Ginzburg-Landau-Abrikosov-Gor'kov theory, $H_{c2}$ is related to the penetration depth $\lambda(T)$ and $H_c(T)$ by

$$H_{c2} = 4\pi\lambda^2(T)H_c^2(T)/\phi_0, \quad (5)$$

where $\phi_0 = 2.07 \times 10^{-7}$ G cm$^2$ is the flux quantum.[65] In the Gorter-Casimir two-fluid model, $\lambda(T) = \lambda(0)(1 - t^4)^{-1/2}$ and $H_c(T) = H_c(0)(1 - t^2)$; thus, $H_{c2}$ should be of the form

$$H_{c2} = \frac{4\pi H_c^2(0)\lambda^2(0)}{\phi_0}\frac{1-t^2}{1+t^2}. \quad (6)$$

As shown in the inset of Fig. 14, the $H_{c2}$ of both compounds follows this relation well, from which $H_{c2}(0)$ and $\lambda(0)$ are determined; $\mu_0 H_{c2}(0) = 1.15$ T and $\lambda(0) = 230$ nm for Cs, and $\mu_0 H_{c2}(0) = 4.81$ T and $\lambda(0) = 220$ nm for Rb. Note that our value of $H_{c2}(0)$ for Rb is much reduced from $H_{c2}(0) = 6$ or 9 T, as reported previously for polycrystalline samples.[8, 31] The $H_{c2}(0)$ of Cs is also slightly smaller than the value of 1.4 T reported for the single crystal used for the dHvA experiments.[34] The similar values of $\lambda(0)$ for Cs and Rb reflect the similarity in the superconducting carrier density on the nearly identical band structures.

The Ginzburg-Landau (GL) coherence length $\xi$ at 0 K is determined to be 17.4 nm for $CsOs_2O_6$ and 8.2 nm for $RbOs_2O_6$ from $H_{c2}(0) = \phi_0 / (2\pi\xi^2)$. The GL parameter $\kappa = \lambda / \xi$ is 14 and 27 for Cs and Rb, respectively; they are typical type-II superconductors. In each compound, $\xi$ is much smaller than the mean free path, clearly indicating that the superconductivity lies at the clean limit.

$\Delta C / T$ for a given magnetic field is related to the slope of $H_{c2}$ through the Maki parameter $\kappa_2$ by the formula

$$\left(\frac{\Delta C}{T}\right) = \frac{1}{4\pi\beta_A(2\kappa_2^2 - 1)}\left(\frac{dH_{c2}}{dT}\right)_T^2, \qquad (7)$$

where $\beta_A = 1.16$ for a triangular vortex lattice.[66, 67] Actually, in the case of $KOs_2O_6$, a hexagonal Abrikosov flux line lattice was observed in recent scanning tunneling spectroscopy measurements.[68] The $T$ dependence of $\kappa_2$ determined by eq. (7) is plotted in Fig. 15. $\kappa_2$ increases with decreasing $T$ below $T_c$, followed by saturation at approximately 21 for Cs, while it seems to further increase to above 50 for Rb. These large enhancements of $\kappa_2$ at low temperatures indicate that the compounds lie close to the pure limit. For example, the theoretical ratio of $\kappa_2$ at $T = 0$ to that at $T_c$ is 2.0 for $\xi / l = 0.05$.[69] In fact, $\xi / l$ for Cs is estimated to be 0.02. The larger enhancement or the tendency toward divergence for Rb may imply that the Rb compound is cleaner than the Cs compound. Furthermore, another Maki parameter $\kappa_1$ is calculated from $H_c$ and $H_{c2}$ by using the equation $H_{c2} = \sqrt{2}\kappa_1 H_c$ and is also plotted in Fig. 15. $\kappa_1$ is almost the same as $\kappa_2$ near $T_c$ and increases more slowly than $\kappa_2$ on cooling.

The $T$ dependences of $\kappa_1$ and $\kappa_2$ theoretically expected for a weak-coupling superconductor at the clean limit[69] are also plotted in Fig. 15. The experimental values are always larger than the calculated ones. This disagreement may arise from neglect of strong-coupling phonons and the effects of FS anisotropy in the theoretical treatment. It is known, however, that the ratio of $\kappa_2$ to $\kappa_1$ depends on the electron mfp; the smaller the ratio $\kappa_2 / \kappa_1$, the smaller the $l$ and, finally, $\kappa_2 \sim \kappa_1$ in the dirty limit.[69] The large values of $\kappa_2 / \kappa_1 > 2$ observed for the present compounds illustrate how clean they are.

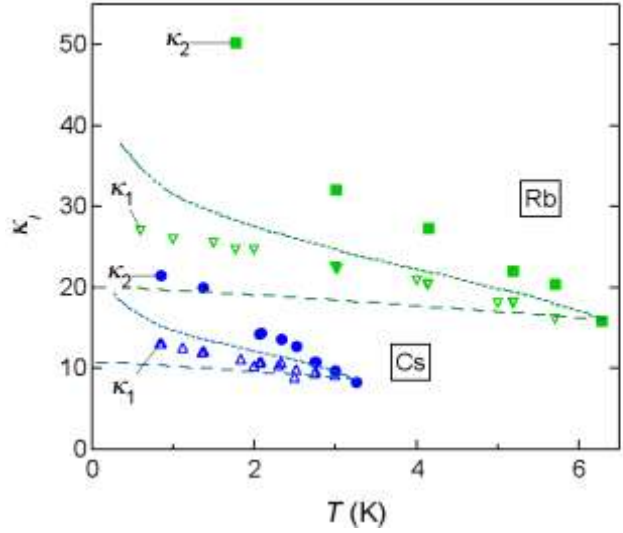

Fig. 15. (Color online) Temperature dependence of the Maki parameters $\kappa_1$ and $\kappa_2$. The broken and dotted curves correspond to those expected theoretically for a weak-coupling superconductor at the clean limit, respectively.[69]

*3.2 Normal-state properties*

*3.2.1 Resistivity and magnetic susceptibility*

The $T$ dependence of resistivity often provides important information about the predominant scatterers of carriers that can be a medium for Cooper pairs in a superconductor. In the case of the $\beta$-pyrochlore oxides, anomalous concave downward resistivity has been commonly observed at high temperatures, followed by $T^2$ behavior at low temperatures for both polycrystalline and single-crystal samples.[21] Dahm and Ueda pointed out that coupling to a local strongly anharmonic phonon mode, which is associated with the rattling of the alkali atoms, is the key for understanding the $T$ dependence; they found $T^2$ and $\sqrt{T}$ behaviors at the low- and high-$T$ limits, respectively.[47]

Figure 16 shows resistivity in a logarithmic plot, measured using high-quality single crystals of all the $\beta$-pyrochlore oxides. It shows $T^2$ behavior at low temperatures and behavior asymptotic to $\sqrt{T}$ at high temperatures for all samples, in good agreement with Dahm and Ueda's expectation.[47] Note that the coefficient of the $T^2$ term increases markedly by more than one order of magnitude from Cs to K, which may indicate a large enhancement in the coupling between the local anharmonic phonons and conduction electrons.

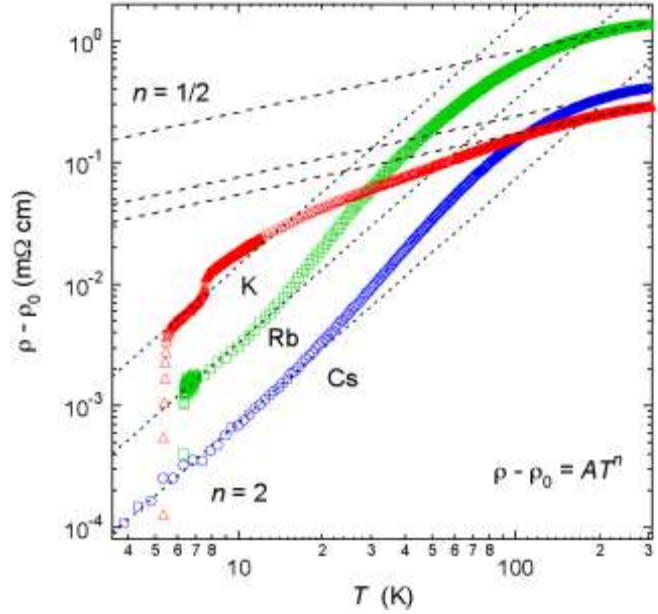

Fig. 16. (Color online) Temperature dependence of resistivity plotted on a logarithmic scale for the three $\beta$-pyrochlore oxides. The residual resistivity $\rho_0$ has been subtracted. All the data were taken at $H = 0$ except for those of K below 10 K, which were recorded at $\mu_0 H = 14$ T. The dotted lines are fits to the $T^2$ form at low temperatures, and the dashed lines are guides to show the $T^{1/2}$ behavior at high temperatures.

The $T^2$ behavior observed for resistivity is reminiscent of electron-electron scattering, which serves as an alternative or additional mechanism for carrier scattering. However, this cannot be the case for the $\beta$-pyrochlore oxides, because there are no corresponding enhancements in magnetic susceptibility and $^{17}$O NMR relaxation rate from Cs to K.[23] Figure 17 shows a comparison of magnetic susceptibility for the three $\beta$-pyrochlore oxides, which shows common $T$-independent Pauli paramagnetism without the signature of Curie-Weiss behavior observed for strongly correlated electron systems such as the related $\alpha$-pyrochlore oxide $Cd_2Os_2O_7$.[70] Moreover, the magnitude of susceptibility is nearly equal among the three compounds with no enhancement toward K. Although Brühwiler $et\ al$. reported a weakly $T$-dependent $\chi$ with a greater magnitude than ours for Rb,[22] we suspect that it was due to coexisting impurities or a low sample quality. The anisotropy in magnetic susceptibility was negligibly small, as reported previously for K.[12]

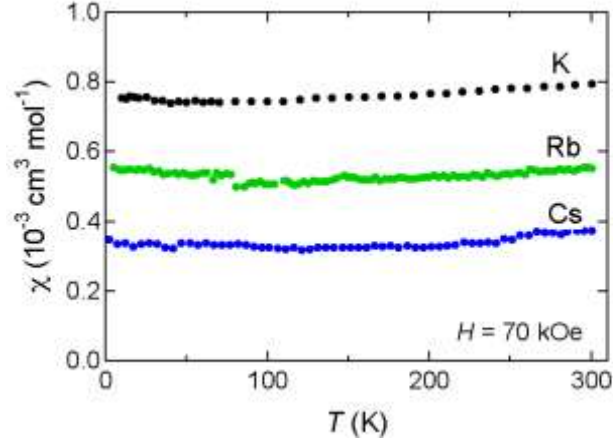

Fig. 17. (Color online) Temperature dependence of magnetic susceptibility $\chi$ measured at $\mu_0 H = 7$ T for all the $\beta$-pyrochlore oxides. For clarity, the sets of data for Rb and K are shifted upward by 2 × $10^{-4}$ and 4 × $10^{-4}$ cm$^3$ mol$^{-1}$, respectively.

*3.2.2 Specific heat*

In order to estimate the characteristic energy of rattling vibrations, we have measured the specific heat over a wide range of $T$ and fitted it following the method adopted for filled skutterudites, Si/Ge clathrates,[43, 71)] and $\beta$-pyrochlore oxides.[12, 15, 21, 22)] It is assumed that the rattler is an Einstein oscillator, because it may be an almost local mode in a cage, and the Os-O framework gives two Debye phonons. Then, the specific heat $C$ per mole is given by

$$C = \gamma T + C_E + a C_{D1} + (8-a) C_{D2}, \qquad (8)$$

where $C_{D1}$ and $C_{D2}$ are the contributions of the Debye phonons, and $C_E$ is that of the Einstein oscillator for the $A$ cation, which takes the form

$$C_E = 3R \left( \frac{\Theta_E}{T} \right)^2 \frac{\exp(\Theta_E/T)}{[\exp(\Theta_E/T)-1]^2}, \qquad (9)$$

where $R$ is the gas constant and $\Theta_E$ is the Einstein temperature. The results of fitting are satisfactory for each compound, as shown in Fig. 18, giving $a = 2.73(5)$, $\Theta_E = 75.1(4)$ K, $\Theta_{D1} = 232(3)$ K, and $\Theta_{D2} = 842(9)$ K for Cs, and $a = 2.39(3)$, $\Theta_E = 66.4(2)$ K, $\Theta_{D1} = 186(2)$ K, and $\Theta_{D2} = 759(6)$ K for Rb. In the previous analysis of KOs$_2$O$_6$, fitting was improved markedly by including another Einstein term, $C_E = b C_{E1} + (1 - b) C_{E2}$, giving $a = 0.24(3)$, $b = 2.55(4)$, $\Theta_{E1} = 22(2)$ K, $\Theta_{E2} = 61(1)$ K, $\Theta_{D1} = 280(3)$ K, and $\Theta_{D2} = 1180(13)$ K.[12)] In the present cases, however, the inclusion of the second Einstein term did not improve the fitting significantly, implying that the single Einstein mode is already an appropriate model of the rattling for Cs and Rb. The evolution of the characteristic energy of the rattling over the series will be discussed later in comparison with other data from spectroscopic experiments and also in relation to the relevant phonons involved in the superconductivity.

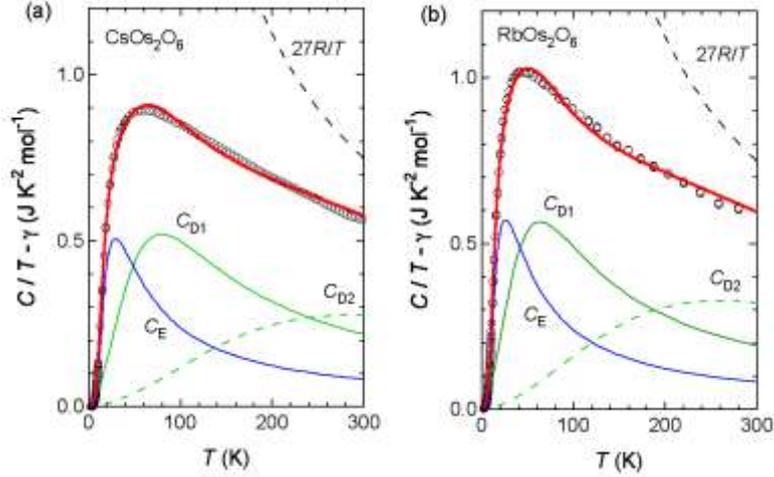

Fig. 18. (Color online) Lattice specific heat over a wide temperature range for (a) Cs and (b) Rb. The data above $T = 10$ K were collected at $H = 0$, while those below 10 K were collected at $\mu_0 H = 2$ T for Cs and at $\mu_0 H = 7$ T for Rb to suppress the superconducting transitions. The values of $\gamma$ have been subtracted from each value of $C/T$. The solid curve on each set of data points corresponds to the fit described in the text, and the lower curves are the separate contributions.

## 4. Discussion

We here discuss the electronic, lattice, and superconducting properties of the $\beta$-pyrochlore oxides on the basis of the various data on $CsOs_2O_6$ and $RbOs_2O_6$ obtained in the present study, in combination with previous data on $KOs_2O_6$ and with reference to those for $Cd_2Re_2O_7$. In order to determine the mechanism of the superconductivity, it is meaningful to investigate the chemical trends of various parameters over the series. We aim to convince the reader that the rattling plays a crucial role in the superconductivity of the $\beta$-pyrochlore oxides by serving as a pairing medium.

*4.1 Basic electronic structures of the β-pyrochlore oxides*

According to first-principles studies, the basic electronic structures of the $\beta$-pyrochlore oxides are identical.[21, 42, 72, 73] The band structure near the Fermi level is given by a manifold of 12 bands arising from Os $5d$ and O $2p$ states with a total bandwidth of ~3 eV. The formal valence of the metals is $A^+$ or $Os^{5.5+}$. The guest alkali metals are always purely ionic, having the related states far above the Fermi level, that is, they donate one electron to the $d$-$p$ bands on the cage to induce metallicity. Both $\alpha$-$Cd_2Re_2O_7$ and $\beta$-$AOs_2O_6$ are semimetals, but they can be discriminated from each other by their carrier density; the former has low carrier density ($4 \times 10^{19}$ cm$^{-3}$),[74] while the latter has many more carriers (~$3 \times 10^{21}$ cm$^{-3}$).[21] The small carrier density in $\alpha$-$Cd_2Re_2O_7$ causes structural instability due to a weak screening effect, resulting in successive structural phase transitions to tetragonal structures.[75] In contrast, the $\beta$-pyrochlore osmates are free from such structural instability.

The calculated FS of $\beta$-$AOs_2O_6$ commonly consists of a pair of closed electron-like sheets around the zone center and a holelike sheet near the zone boundary.[21, 42, 73] Note the characteristic shape of the electron-like FS: the two sheets are nearly homothetic with an octahedral shape, which may give rise to instability due to the FS nesting. Moreover, the sheets are nearly parallel to each other for K, while the outer sheet is dented toward the inner sheet for Cs. Very recent dHvA experiments carried out on $CsOs_2O_6$ confirmed the results of the band structure calculations except for the fact that the dents become small through holes in the <111> direction, indicating that the van Hove singularity is located slightly above the Fermi level.[34] The observed FS has been reproduced successfully by incorporating a slight rigid band shift.[34] Further important information from the dHvA experiments are effective mass enhancements by a factor of 3 - 4, consistent with the present specific heat data, and a large mfp of 700 - 1000 nm depending on the band branches.

Table I. Normal-state electronic parameters for $\alpha$-pyrochlore $Cd_2Re_2O_7$ and three $\beta$-pyrochlore oxides $AOs_2O_6$.

| | $\alpha$-$Cd_2Re_2O_7$[2,6] | $\beta$-$CsOs_2O_6$ | $\beta$-$RbOs_2O_6$ | $\beta$-$KOs_2O_6$[12] |
|---|---|---|---|---|
| $\gamma_{exp}$ (mJ $K^{-2}$ $mol^{-1}$) | 30.2 | 41.4 | 44.7 | 70 |
| $\gamma_{band}$ (mJ $K^{-2}$ $mol^{-1}$)[a] | 11.5 | 11.0 | 10.2 | 9.6 |
| $\gamma_{exp}$ / $\gamma_{band}$ | 2.63 | 3.76 | 4.38 | 7.3 |
| $\chi^{exp}$ ($10^{-4}$ $cm^3$ $mol^{-1}$) | 3.0 | 3.3 | 3.1 | 3.5 |
| $\chi_P^{exp}$ ($10^{-4}$ $cm^3$ $mol^{-1}$) | 1.5 | 2.0 | 1.6 | 1.6 |
| $\chi_P^{band}$ ($10^{-4}$ $cm^3$ $mol^{-1}$) | 1.6 | 1.5 | 1.4 | 1.3 |
| $\chi_P^{exp}$ / $\chi_P^{band}$ | 0.94 | 1.3 | 1.1 | 1.2 |
| $R_W$ | 0.50 | 0.48 | 0.36 | 0.14 |
| $A$ ($\mu\Omega$ $K^{-2}$ $cm^{-1}$) | | 0.0074 | 0.0321 | 0.142 |
| $A$ / $\gamma^2$ ($10^{-5}$) | | 0.43 | 1.6 | 2.9 |
| $l$ [$\rho_0$] (nm) | 700 | 850 | 580 | 640 |

[a]These $\gamma_{band}$ values for the $\beta$-pyrochlore oxides[21] are slightly smaller than those given by Saniz and Freeman,[73] which were attained by using calculated lattice constants that are larger than the experimental ones.

In the absence of contributions from the alkali metals, the density of states (DOS) at the Fermi level is roughly proportional to the lattice constant; the smaller the lattice constant from Cs to K, the smaller the calculated DOS. This tendency is opposite to our experimental observations. The experimental Sommerfeld coefficient $\gamma_{exp}$, which is proportional to DOS, is compared with the calculated $\gamma_{band}$ in Table I and Fig. 19(a). The enhancement factor $\gamma_{exp}$ / $\gamma_{band}$ is 2.63 for $Cd_2Re_2O_7$ and 3.76, 4.38, and 7.3 for Cs, Rb, and $KOs_2O_6$, respectively. In contrast, there are no corresponding enhancements in magnetic susceptibility, as already shown in Fig. 17; $\chi_P^{exp}$ is always approximately equal to $\chi_P^{band}$ with a slight variation, as shown in Fig. 19(b), where $\chi_P^{band} = 2\mu_B^2 N(0) \sim 1.6 \times 10^{-4}$ $cm^3$ $mol^{-1}$. Thus, the Wilson ratio,

$$R_W = (\pi^2/3)(k_B/\mu_0)^2 (\chi_P/\gamma) = 72.949 (\chi_P/\gamma), \quad (10)$$

is smaller than 0.5 and is particularly small for K, $R_W = 0.14$. This clearly indicates that the large mass enhancement is due to strong e-ph interactions and that the e-e interactions would have a minor effect, even if they exist. Moreover, note that, as $\gamma$ increases from Cs to K, the coefficient $A$ of the $T^2$ term in $\rho$ increases, as shown in Figs. 16 and 19(c), suggesting a common origin for these enhancements. The values of the Kadowaki-Woods ratio $A/\gamma^2$ are 0.43, 1.6, and 2.9 for Cs, Rb, and K, respectively, which are close to the universal value of $10^{-5}$ expected for strongly correlated electron systems, but this may be a coincidence. The $T^2$ term in $\rho$ should be ascribed to the strong scattering by low-energy and strongly anharmonic phonons, as pointed out in previous studies.[12, 21, 29, 47]

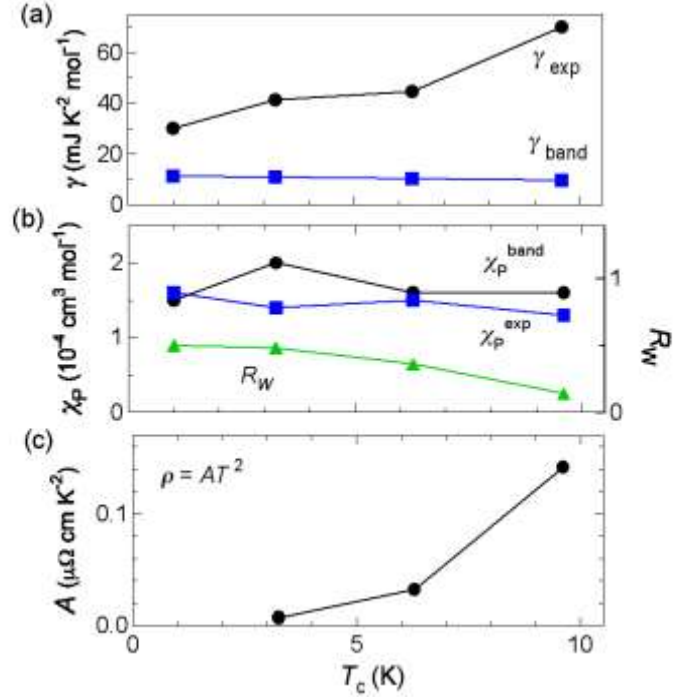

Fig. 19. (Color online) Evolution of normal-state properties as a function of $T_c$ over the series of pyrochlore oxides: (a) Sommerfeld coefficient $\gamma$, (b) Pauli paramagnetic susceptibility $\chi_P$, and (c) the coefficient $A$ of the $T^2$ term in resistivity. The Wilson ratio $R_W$ is also plotted in (b). The experimental values for $\gamma$ and $\chi_P$ are compared with values obtained from band structure calculations in (a) and (b).

Further evidence for strong $e$-ph interactions is found in the large values of saturated resistivity at high temperatures, which are interpreted to correspond to the Ioffe-Regel limit: the mfp becomes of the order of the interatomic distance of ~0.36 nm. As estimated from the values of $\rho_0$, in contrast, the mfp at $T = 0$, as listed in Table I, is large enough for extrinsic scattering due to impurities or crystalline defects to be negligibly small. A similar crossover from $T^2$ behavior to saturated behavior with increasing temperature has been observed in the A-15 compounds such as $Nb_3Sn$,[76, 77] which are typical strong-coupling superconductors with a maximum $T_c$ of 23 K. The origin of this crossover has been ascribed to scattering by a soft phonon mode with nonnegligible anharmonicity, which may also cause the "high-$T_c$" superconductivity in the strong-coupling regime. We conclude that there must be certain similarities between the A-15 compounds and the $\beta$-pyrochlore superconductors.

*4.2 Lattice properties*

As mentioned in the introduction, a characteristic of the *β*-pyrochlore structure is that the *A* cation is weakly bound to the cage made of $OsO_6$ octahedra and can "rattle" in an anharmonic potential (Fig. 1), particularly in the case of the smallest K cation.[35, 42, 78-80] Table II summarizes the crystallographic parameters of the pyrochlore oxides obtained in previous structural studies.[35, 36, 81] Since the lattice parameter *a* and, thus, the size of the cage shrink only slightly from Cs to K, the atomic excursion of the *A* cation inside the cage increases enormously with decreasing ionic radius (0.167, 0.152, and 0.138 nm for $Cs^+$, $Rb^+$, and $K^+$ ions coordinated with six oxygen ions, respectively), as evidenced by a large increase in the isotropic atomic displacement parameter $U_{iso}$. Note that the value of $U_{iso}$ of $7.7 \times 10^{-4}$ $nm^2$ for K is exceptionally large for such a heavy metal atom. In terms of lattice dynamics, it is expected that the rattling vibration appears as a low-lying local mode, that is, a dispersionless mode, and the cage gives ordinary acoustic and optical modes at higher energies. The former rattling mode should manifest itself as an Einstein mode in the specific heat or a sharp peak in the phonon DOS, which was actually observed in recent INS experiments.[45, 46] The energy of the rattling modes determined thus far by various methods is listed in Table II.

Table II. Crystallographic and phonon data at room temperature for *α*-pyrochlore $Cd_2Re_2O_7$ and three *β*-pyrochlore oxides $AOs_2O_6$.

|  | α-$Cd_2Re_2O_7$ | β-$CsOs_2O_6$ | β-$RbOs_2O_6$ | β-$KOs_2O_6$ |
|---|---|---|---|---|
| *a* (nm)[a] | 1.0219[82] | 1.01525[81] | 1.011393[36] | 1.0089[35] |
| *x*(O)[b] | 0.3089 | 0.3146 | 0.3168 | 0.3145 |
| $U_{iso}$ ($10^{-4}$ $nm^2$)[c] | 1.1 | 2.5 | 3.41 | 7.4 |
| $\Theta_E$ (K) |  | 75.1 | 66.4 | 22 / 61 |
| $\Theta_D$ (K) | 458 | 232 / 842 | 186 / 759 | 280 / 1180 |
| $E(T_{2g})$ [Raman] (meV)[h] |  | 7.6 | 7.4, 7.5 | 8.8, 9.4 |
| $E(T_{2g})$ [INS] (meV)[k] |  | 7.2 | 7.3 | 6.8 |
| $E(T_{1u})$ [INS] (meV)[k] |  | 5.9 | 5.4 | 5.5 |

[a]lattice constant: space group *Fd-3m*
[b]atomic coordinate of the 48*f* oxygen: (*x* 0.125 0.125)
[c]isotropic atomic displacement parameter at room temperature for Cd or *A* cations
[h]phonon energy at room temperature[40, 41, 83]
[k]phonon energy at room temperature[46]

According to the mode analysis used to assign observed peaks in Raman spectroscopy,[40, 41, 49] there are two modes attributable to the vibrations of the *A* atoms in the *β*-pyrochlore structure: $T_{2g}$ and $T_{1u}$. The $T_{2g}$ mode is Raman active and was actually observed in the Raman spectra at 8.8 meV[40, 49] or 9.4 meV[41] for K. Smaller values of ~7.5 meV were found for Cs and Rb,[41, 83] as listed in Table II. On the other hand, in the first INS experiments by Sasai *et al*. at room temperature using polycrystalline samples of all the *β*-pyrochlore oxides, a broad peak was found at 6.5 meV.[45] Later in the high-resolution INS experiments by Mutka *et al*., a double sharp peak instead of a single broad peak was detected for each compound.[46] Figure 20 shows the chemical trend of the energy of these rattling modes as a function of $T_c$. The $T_{2g}$ mode from the Raman data appears to correspond to the higher-energy peak in the INS spectra for Cs and Rb, which means that the lower-energy peak is ascribed to the missing Raman inactive $T_{1u}$ mode. On the other hand, the value of $k_B\Theta_E$ obtained from specific heat measurements coincides with the energy of the $T_{1u}$ mode for each compound. This is reasonable, because a thermal excitation to the lowest energy level should be detected through specific heat measurements.

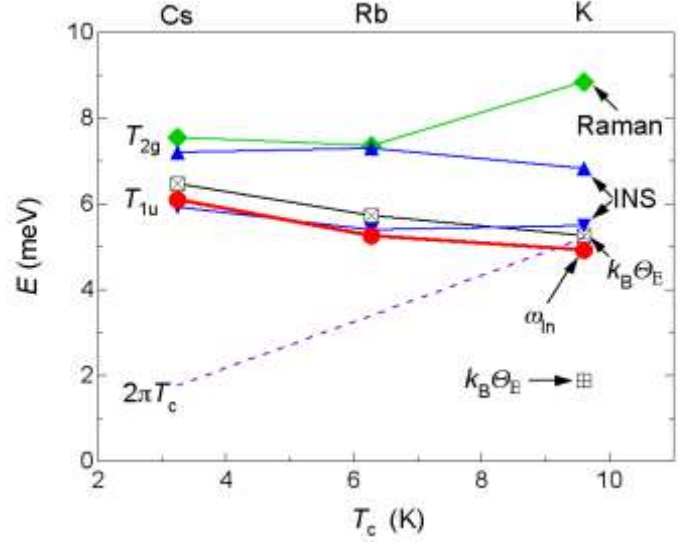

Fig. 20. (Color online) Evolution of the characteristic energy for the low-lying phonon modes as a function of $T_c$ over the series of $\beta$-pyrochlore oxides. The energy of the Einstein modes from the present specific heat measurements $k_B\Theta_E$ is compared with those of the two modes, $T_{2g}$ and $T_{1u}$, determined at room temperature by Raman spectroscopy[40, 83] and INS experiments.[46] The average phonon frequency responsible for the occurrence of superconductivity $\omega_{\ln}$ is also plotted.

Since the two peaks in the INS spectra are sharp for Cs and Rb, their vibrations must be local modes expected from the simple rattling picture. In contrast, the peaks are relatively broad for K.[46] Moreover, there is a significantly large difference in the energy of the $T_{2g}$ mode between the Raman and INS data for K (see Fig. 20). It is likely that the rattling of the K atom is not simply a local vibration; it interacts significantly with its surroundings or the nearest-neighbor K ions. In addition, a large shift of the peaks to lower energy was found on cooling, particularly for K, in both the Raman and INS experiments: the energy of the K $T_{2g}$ mode is reduced from 8.8 meV at room temperature to 7.8 meV at 4 K according to the Raman data,[83] while a greater softening down to 3.4 meV was detected for the K $T_{1u}$ mode according to the INS data.[46] One of the two Einstein modes with a lower energy of 2 meV found in previous specific heat measurements of K[12, 22] is thought to correspond to this soft mode. In previous PES experiments at a low temperature the corresponding low-energy mode at 2.1 meV was also found.[18]

*4.3 Superconducting properties*

Here we summarize the superconducting properties of $CsOs_2O_6$ and $RbOs_2O_6$ and compare them with those of $KOs_2O_6$ and $Cd_2Re_2O_7$. All the thermodynamic parameters deduced in the present study are listed in Table III. Only the lower critical field $H_{c1}$ has not been determined experimentally, and it is estimated from the equation

$$H_{c1}(0) = \frac{\varphi_0}{4\pi\lambda^2}\left[\ln\left(\frac{\lambda}{\xi}\right) + 0.08\right] \quad (11)$$

to be 8.2 and 11.1 mT for Cs and Rb, respectively.

Table III also compares our data for $RbOs_2O_6$ with the previous data given by Brühwiler et al.[31] Most parameters are in good agreement; in particular, the values of $H_c(0)$ are nearly equal. Some notable differences originate from the different values of $H_{c2}$: $\mu_0 H_{c2}(0) = 4.81$ T in the present study is smaller than the value of 6 T obtained by Brühwiler et al. Since the value of $H_{c2}$ is sensitive to the mfp and is increased by impurity scattering, the present results give more intrinsic parameters in a regime closer to the clean limit.

Table III. Thermodynamic parameters of superconductivity for $\alpha$-pyrochlore $Cd_2Re_2O_7$ and three $\beta$-pyrochlore oxides $AOs_2O_6$.

|  | $\alpha$-$Cd_2Re_2O_7$[6] | $\beta$-$CsOs_2O_6$ | $\beta$-$RbOs_2O_6$ | $\beta$-$RbOs_2O_6$[31] | $\beta$-$KOs_2O_6$[12] |
|---|---|---|---|---|---|
| $T_c$ (K) | 0.97 | 3.25 | 6.28 | 6.4 | 9.6 |
| $\mu_0 H_{c2}(0)$ (T) | 0.29 | 1.15 | 4.81 | 6 | 30.6 |
| $\mu_0 H_c(0)$ (mT) | 15 | 59.0 | 125.9 | 124.9 | 255 |
| $\mu_0 H_{c1}(0)$ (mT) | 2.1 | 8.2 | 11.1 | 9.2 | 10.1 |
| $-d\mu_0 H_{c2}/dT|_{T_c}$ (T K$^{-1}$) | 0.42 | 0.44 | 0.88 | 1.2 | 3.61 |
| $-d\mu_0 H_c/dT|_{T_c}$ (mT K$^{-1}$) | 24.7 | 30.3 | 36.5 | 37.0 | 56.1 |
| $\xi$ (nm) | 34 | 17 | 8.3 | 7.4 | 3.3 |
| $\lambda$ (nm) | 460 | 230 | 220 | 252 | 270 |
| $\kappa$ | 14 | 14 | 27 | 34 | 82 |
| $\Delta C / \gamma T_c$ | 1.15 | 1.49 | 1.83 | 1.9 | 2.87 |
| $\gamma T_c^2 / H_c(0)^2$ | 0.157 | 0.160 | 0.142 | 0.15 | 0.128 |
| $l / \xi$ | 21 | 50 | 70 |  | 120 |
| $2\Delta(0) / k_B T_c$ |  |  |  |  |  |
| $\alpha$ model: single gap |  | 3.66 | 4.00 |  | 5.00 |
| $\alpha$ model: two gaps |  | 2.4 / 3.8 | 3.2 / 4.8 |  | - |
| Toxen relation |  | 3.33 | 3.64 |  | 4.00 |
| Low-$T$ fitting |  | 2.73 | 2.78 |  | 4.69 |
| Strong-coupling correction [$H_c(0)$] |  | 3.75 | 4.27 |  | 4.89 |

Figure 21 shows the superconducting transitions with changes in specific heat for all the $\beta$-pyrochlore oxides and $\alpha$-$Cd_2Re_2O_7$. It is apparent that the jump at $T_c$ increases enormously with increasing $T_c$ and that $\gamma$ increases accordingly. The chemical trends of the superconducting properties over the series are shown in Fig. 22, where $\Delta C / \gamma T_c$ and $2\Delta / k_B T_c$, determined by fitting to the $\alpha$ model with a single-gap parameter, and $H_{c2}$ or $\xi$ are plotted as functions of $T_c$. The character of superconductivity changes systematically from weak BCS-type coupling to strong coupling with increasing $T_c$. Since $\xi$ decreases in the same direction, the size of Cooper pairs decreases with increasing $T_c$, implying an enhanced pair potential.

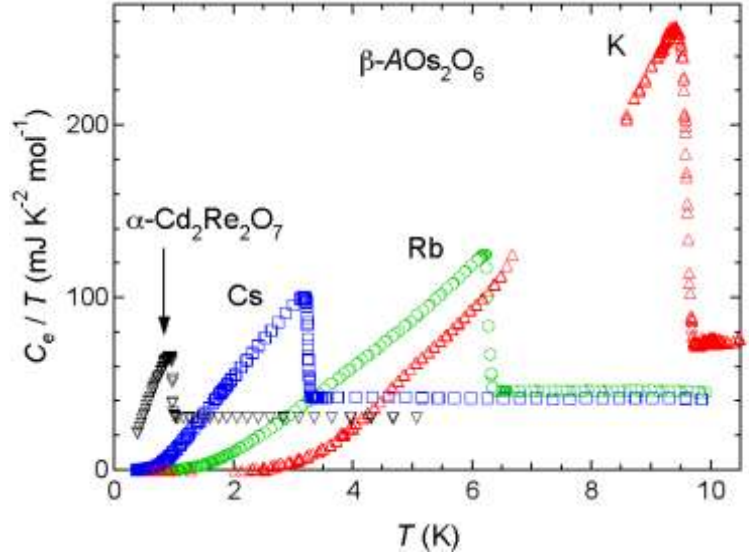

Fig. 21. (Color online)   Superconducting transitions observed in specific heat for $\alpha$-Cd$_2$Re$_2$O$_7$ and the three $\beta$-pyrochlore oxides.

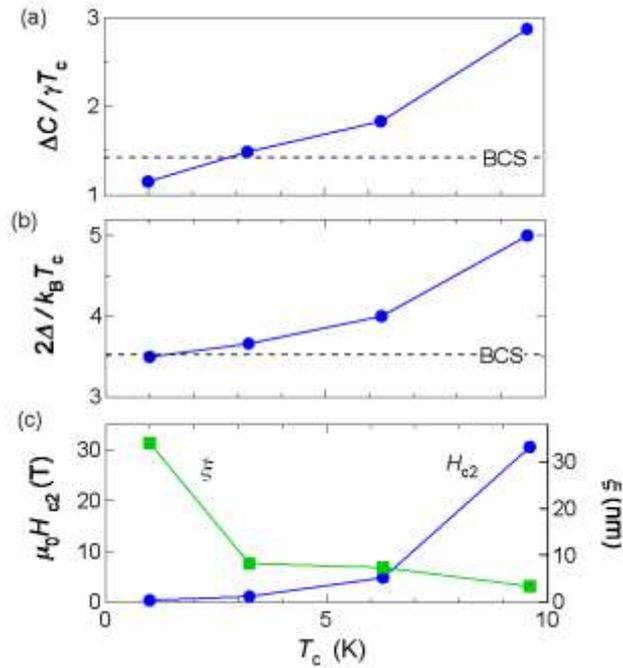

Fig. 22. (Color online)   Evolution of superconducting parameters as a function of $T_c$.  (a) Jump in specific heat at $T_c$, (b) superconducting gap determined by fitting to the single-gap $\alpha$ model, and (c) upper critical field $H_{c2}$ and superconducting coherence length $\xi$.

*4.4 Strong-coupling correction analyses*

To investigate *e*-ph coupling in a superconductor, of primary importance is the *e*-ph interaction function $\alpha^2(\omega)F(\omega)$, where $\alpha^2(\omega)$ is the coupling function and $F(\omega)$ is the phonon DOS. $\alpha^2(\omega)$ is usually extracted from tunneling experiments. However, because they have not yet been carried out on the $\beta$ pyrochlores, we deduce the logarithmically averaged mean phonon frequency $\omega_{\ln}$, which is related to $\alpha^2(\omega)F(\omega)$ by

$$\omega_{\ln} = \exp\left(\frac{2}{\lambda}\int_0^\infty d\omega \frac{\alpha^2(\omega)F(\omega)}{\omega}\ln\omega\right), \quad (12)$$

$$\lambda = 2\int_0^\infty d\omega \frac{\alpha^2(\omega)F(\omega)}{\omega}. \quad (13)$$

$\omega_{\ln}$ can be estimated on the basis of the strong-coupling correction analyses developed by Marsiglio and Carbotte.[84] Approximate analytic formulae with simple correction factors of the BCS values are given as a function of the single parameter $x = T_c / \omega_{\ln}$:

$$\frac{\Delta C(T_c)}{\gamma T_c} = 1.43\left[1 + 53x^2 \ln\left(\frac{1}{3x}\right)\right], \quad (14)$$

$$\frac{\gamma T_c^2}{H_c^2(0)} = 0.168\left[1 - 12.2x^2 \ln\left(\frac{1}{3x}\right)\right], \quad (15)$$

$$\frac{2\Delta(0)}{k_B T_c} = 3.53\left[1 + 12.5x^2 \ln\left(\frac{1}{2x}\right)\right]. \quad (16)$$

Equations (14) and (15) are shown in Fig. 23, where various materials used to obtain these empirical formulae are plotted. We estimate the values of $x$ for the Cs and Rb compounds by these two equations using the corresponding experimental values: $x = 0.016$ (0.054) from $\Delta C(T_c) / \gamma T_c = 1.49$ (1.83) and $x = 0.046$ (0.103) from $\gamma T_c^2 / H_c^2(0) = 0.160$ (0.142) for Cs (Rb). Remarkably, there are large differences between the two values of $x$ determined from $\Delta C$ and $H_c$ for each compound, as compared in Fig. 23 and Table IV. Empirically, the two evaluations should give similar values of $x$. In the case of Nb, for example, $x = 0.064$ and 0.068 from $\Delta C$ and $H_c$, respectively.[84] In addition, the two evaluations give $x = 0.094$ and 0.092 for a Chevrel phase compound $Mo_6Se_8$ having similar superconducting parameters to Rb.[85] For $KOs_2O_6$, the two values of $x$ are nearly equal: 0.163 and 0.168.[12] Thus, the present disagreements for Cs and Rb are exceptional and can be related to the anisotropy of the gap; the two evaluations can give different results for an anisotropic gap, because $\Delta C(T_c)$ is the quantity at high temperature, while $H_c(0)$ is the quantity at $T = 0$. We take the latter as more appropriate for deducing values of $x$, because $H_c(0)$ is directly related to the condensation energy of the superconducting state. Thus, we have determined $x$ and obtained $\omega_{\ln}$ as $\omega_{\ln} = 70.7$ and 61.0 K for Cs and Rb, respectively. The value of $\omega_{\ln}$ for Rb is smaller than a previously reported value of 107 K deduced from the strong-coupling correction of $\Delta C$.[31]

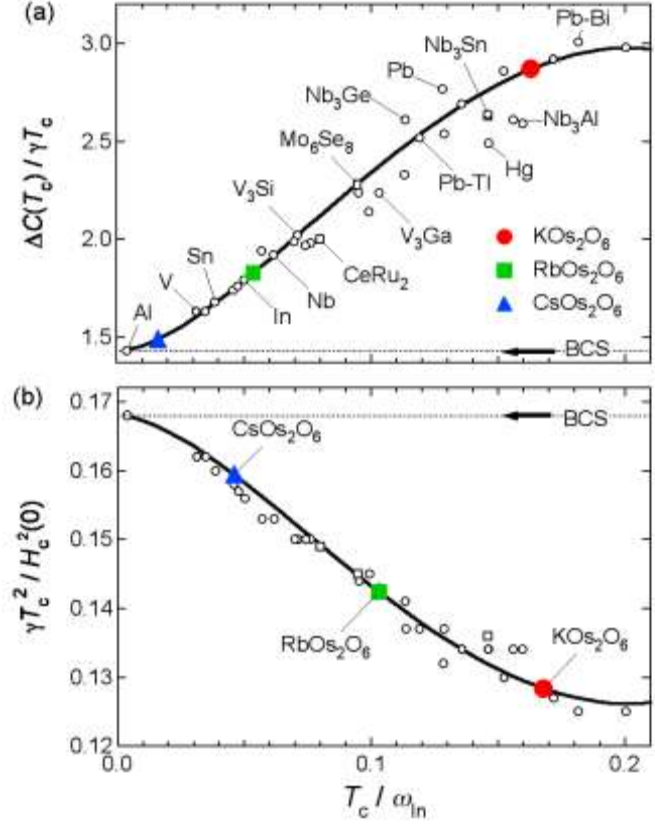

Fig. 23. (Color online)  (a) Normalized jump in specific heat at $T_c$, $\Delta C(T_c) / \gamma T_c$, and (b) the ratio $\gamma T_c^2 / H_c^2(0)$ as a function of $T_c / \omega_{\ln}$.  Materials shown by open circles, in increasing order of $T_c / \omega_{\ln}$, are Al, V, Ta, Sn, Tl, $Tl_{0.9}Bi_{0.1}$, In, Nb, Nb, $V_3Si$, $V_3Si$, Nb, Mo, $Pb_{0.4}Tl_{0.6}$, La, $V_3Ga$, $Nb_3Al$, $Nb_3Ge$, $Pb_{0.6}Tl_{0.4}$, $Nb_3Al$, Pb, $Pb_{0.8}Tl_{0.2}$, Hg, $Nb_3Sn$, $Pb_{0.9}Bi_{0.1}$, $Nb_3Al$, $Nb_3Ge$, $Pb_{0.8}Bi_{0.2}$, $Pb_{0.7}Bi_{0.3}$, and $Pb_{0.65}Bi_{0.35}$, which are from ref. 84. Those marked by open squares are $CeRu_2$[62] and $Mo_6Se_8$,[85] in the same order.  The $\beta$-pyrochlore oxides are shown by solid symbols: Cs (triangle), Rb (square), and K (circle).  The solid curves in (a) and (b) correspond to eqs. (14) and (15), respectively.

*4.5 Possible anisotropy in the superconducting gap*

Here we discuss on the possibility of anisotropy in the superconducting gap or multigaps on different FSs.  There are several indications of anisotropy for the Cs and Rb compounds: 1) the specific heat data below $T_c$ can be fitted reasonably well by the $\alpha$ model assuming two gaps, as shown in Fig. 9; 2) the magnitude of the gap determined at low temperatures from the specific heat is significantly smaller than that determined by the single-gap $\alpha$-model fitting, as shown in Fig. 8 and also in Fig. 24, where $C_{es} / \gamma T_c$ is plotted against $T_c / T$; 3) the nonlinear field dependence of $\gamma(H)$ shown in Fig. 10 suggests that there is a region on the FS with a smaller gap; 4) the $H_c$ deviation function exhibits an unusual $T$ dependence for Rb, as shown in Fig. 12, and 5) the strong-coupling correction analyses based on the two thermodynamic parameters, $\Delta C / \gamma T_c$ and $\gamma T_c^2 / H_c^2(0)$, gave inconsistent values of $\omega_{\ln}$.  Table III lists the magnitude of the superconducting gap deduced by various methods.  All these facts strongly indicate that there is anisotropy in the superconducting gap for Cs and Rb or that a multiple gap exists. We cannot determine which of these possibilities is true in the present study. In the latter case, however, one would expect two or more gaps with a similar weight ratio for all the $\beta$-pyrochlore oxides, taking into account the similarity of the band structures. In this case, it is likely that the two gaps can be ascribed to the electron- and holelike FSs with a $\gamma$ ratio of approximately 2:1 found in the band structure calculations.[86] In reality, however, the $\gamma$ ratio for the two bands found for Cs and Rb is considerably different from this value: $\gamma_1 / (\gamma_1 + \gamma_2) = 0.11$ and 0.51, respectively. Therefore, we consider it more realistic to assume the anisotropy of the gap instead of the existence of a multigap.

The $k$-space anisotropy of the conventional $s$-wave gap is given by $\Delta = <\Delta>[1 + a(\Omega)]$, where $\Omega$ is the solid angle in $k$ space, $<\Delta>$ is the mean gap value over all orientations, and $a(\Omega)$ is the anisotropy function satisfying the condition $<a(\Omega)> = 0$.[87] It is known that the effect of anisotropy on the $T$ dependences of the gap parameter, the critical field, and the specific heat near $T_c$ is small for most superconductors and proportional to the mean-square anisotropy $<a^2>$, which is on the order of 0.02 for a typical superconductor.[87] In contrast, it manifests itself more pronouncedly in the low-temperature specific heat or the nuclear spin-lattice relaxation rate.

For RbOs$_2$O$_6$, Manalo et al. analyzed the $H_c(T)$ and $H_{c2}(T)$ data within the framework of $s$-wave Eliashberg theory and obtained a small gap anisotropy parameter of $<a^2> \sim 0.005$ and a Fermi velocity anisotropy parameter of $<b^2> = 0.3 - 0.35$.[63] In contrast, Magishi et al. obtained a larger anisotropy of the $s$-wave gap based on the results of Rb NMR measurements at low temperature.[14] The anisotropy renormalizes the gap at the low-temperature limit as $2\Delta / k_B T_c = (2\Delta / k_B T_c)^0 [1 - (3/2)<a^2>]$, where $(2\Delta / k_B T_c)^0$ is the gap size in the absence of anisotropy.[87] They found in their analyses of the Rb NMR relaxation rate that $(2\Delta / k_B T_c)^0 = 5.1$ and $<a^2> = 0.25$.[14]

In our analyses on the specific heat data of Rb by the $\alpha$ model, we obtained $2\Delta / k_B T_c = 4.00$ by assuming a single gap and $2\Delta / k_B T_c = 3.2$ and 4.8 with a nearly equal weight for two gaps. Assuming that these gap values approximate the lower and upper bounds for the anisotropy, one obtains $(2\Delta / k_B T_c)^0 = 4.0$ and $<a^2> = 0.04$. This gap value is equal to the gap size estimated from fitting by the single-gap $\alpha$ model and close to that obtained from the strong-coupling analyses (4.27). The mean-square anisotropy is in between the above values and similar to that of Nb, $<a^2> = 0.034$.[88] On the other hand, it is difficult to estimate $<a^2>$ in the same way for Cs, because the two-gap fitting gave unsatisfactory results; the contribution from the smaller gap is too small.

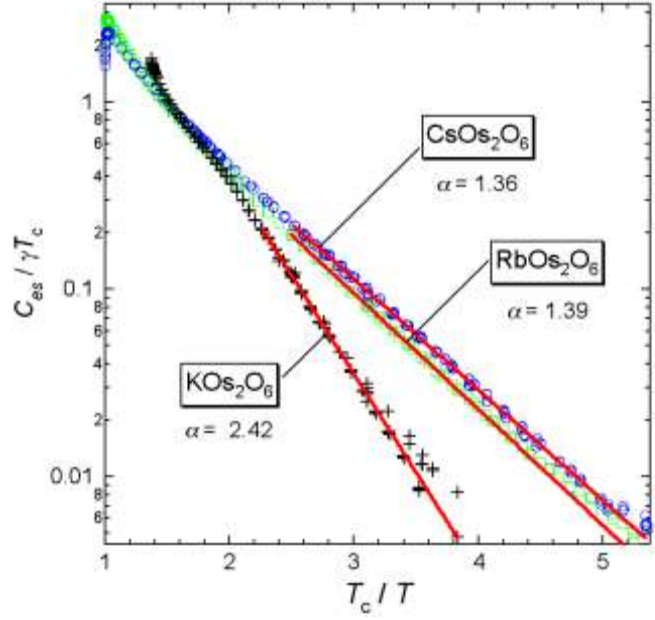

Fig. 24. (Color online) Electronic specific heat divided by $\gamma T_c$ as a function of inverse reduced temperature for the three $\beta$-pyrochlore oxides. The solid line on each data set is a fit to the exponential form.

Alternatively, we estimate $\langle a^2 \rangle$ on the basis of the results of the strong-coupling correction analyses. As mentioned in the last section, we have obtained $x = 0.046$ (0.103) for Cs (Rb) from the $H_c(0)$ using eq. (15). Using these values of $x$, one can estimate the jump in specific heat at $T_c$ and the gap size in the absence of anisotropy from eqs. (14) and (16): $[\Delta C(T_c) / \gamma T_c]^0 = 1.75$ (2.37) and $[2\Delta(0) / k_B T_c]^0 = 3.75$ (4.27) for Cs (Rb). The former values are much larger than the experimental values of 1.49 (1.83). It is known that $\Delta C(T_c) / \gamma T_c$ is renormalized by a factor of $(1 - 2.351\langle a^2 \rangle)$ in the presence of anisotropy for a weak-coupling superconductor.[87] Simply applying this factor, one obtains $\langle a^2 \rangle = 0.040$ (0.068) for Cs (Rb). On the other hand, a comparison between the above value of $[2\Delta(0) / k_B T_c]^0 = 3.75$ (4.27) and $2\Delta(0) / k_B T_c = 2.73$ (2.78) obtained from the low-temperature specific heat yields $\langle a^2 \rangle = 0.18$ (0.23), applying the renormalization factor of $[1 - (3/2)\langle a^2 \rangle]$. The large differences in $\langle a^2 \rangle$ from the two evaluations is thought to originate from the simple assumption of a spherical FS. In order to determine true three-dimensional anisotropy, we have to consider the actual shape of the FSs.

It is worth noting that the lowest-$T$ gap size is nearly the same for Cs and Rb, as compared in Fig. 24. This suggests that in both Cs and Rb there is an area at a certain part of the FSs having a small gap compared with the other part. This means, in turn, that the differences between their superconducting properties should originate from the other part of the FS with large gaps, where the $e$-ph coupling must have a greater effect than that in the low-gap part, resulting in a larger enhancement for Rb than Cs.

On the other hand, in the case of $KOs_2O_6$, the gap size obtained from the specific heat at the low-temperature limit is large, $2\alpha = 4.69$, and comparable to that obtained from fitting to the single-gap $\alpha$ model, $2\alpha = 5.00$. Moreover, the strong-coupling correction analyses based on $\Delta C(T_c)$ and $H_c(0)$ gave nearly the same values of $x$. The gap value obtained from $x$ using eq. (14) is 4.89, also close to $2\alpha = 5.00$. Thus, it seems that there is no critical anisotropy in the gap. However, as shown in Fig. 21, the $C_e/T$ of K exhibits a strange $T$ dependence at approximately $T_c/2$. Moreover, a previous $\mu$SR study by Koda et al. found a strong magnetic-field dependence of penetration depth and concluded the existence of anisotropic s-wave pairing or multigaps with a very small minimum gap of less than 0.2 meV.[19] In addition, very recent scanning tunneling spectroscopy measurements on a $KOs_2O_6$ single crystal revealed fully gapped superconductivity with a larger gap anisotropy of 30%.[68] Although all the results are consistent with each other regarding the anisotropy of s-wave superconductivity, the magnitude of anisotropy in the gap remains a controversial issue. Consequently, there is always considerable anisotropy in the superconducting gap of the $\beta$-pyrochlore oxides, which must reflect a certain aspect of e-ph interactions that is responsible for the superconductivity.

*4.6 Relevant phonons*

We have successfully obtained reliable values for the mean phonon frequency $\omega_{ln}$ for all the $\beta$-pyrochlore oxides, as listed in Table IV. They are 70.7, 61.0, and 57.1 K for Cs, Rb, and K, respectively, which gradually decrease with decreasing ionic size of the alkali metals. These values are compared in Fig. 20 with the characteristic energy of low-energy phonons estimated from specific heat measurements in the present study and from previous spectroscopic experiments, as mentioned before. The corresponding Einstein temperature $\Theta_E$ decreases to 75.1, 66.4, and 61 K, always several kelvin above $\omega_{ln}$. The energy of the $T_{1u}$ mode observed in the phonon DOS from INS experiments varies almost in synchronization with $\Theta_E$ and $\omega_{ln}$. Therefore, it is obvious that this low-lying mode dominates $\alpha^2(\omega)F(\omega)$ and thus mediates the Cooper pairing in the $\beta$-pyrochlore oxides. Since the rattling of the alkali cations is clearly the entity of this low-lying mode, we conclude that the superconductivity is actually caused by the rattling. It may be important to emphasize that $\omega_{ln}$ coincides with the energy of the $T_{1u}$ mode, not the $T_{2g}$ mode, suggesting that the electron-rattler interactions associated with the $T_{1u}$ mode are crucial. Also note that the reason why we can so definitely conclude the occurrence of rattling-induced superconductivity based on the simple analyses of $\omega_{ln}$ is that the rattling is an almost dispersionless mode with a sharp peak at a specific energy in the phonon DOS, which is not the case for most superconductors with relevant dispersive phonons.

It has been predicted theoretically that the most effective range of $\alpha^2F(\omega)$ to enhance $T_c$ lies slightly above $2\pi T_c$.[89] As shown in Fig. 20, the line of $2\pi T_c$ lies below $\omega_{ln}$ for Cs and Rb, and coincides with $\omega_{ln}$ at K. This trend gives a reasonable explanation for the increases of $T_c$ and the pair potential from Cs to K.

In the case of $KOs_2O_6$, an additional lower energy mode exists at ~20 K, which was found in specific heat, PES,[18] and NMR measurements.[23] In recent INS experiments a lower-energy mode was found to appear at approximately 3 meV at low temperatures.[46] This mode exhibits a large softening and is thought to be related to the large anharmonicity of the K vibration. However, the determined value of $\omega_{\ln}$ of 57.1 K (~5 meV) implies that this mode does not affect the occurrence of superconductivity, because the energy is too low, much lower than the superconducting gap of $2\Delta \sim 50$ K, to contribute to the pairing. It has also been pointed out that the first-order phase transition at $T_p = 7.5$ K below $T_c$ may be related to the softening of this lowest energy mode and thus does not strongly affect the superconductivity itself; this transition is only accompanied by a change in the mfp.[12] The superconductivity of $KOs_2O_6$ must therefore take place via a virtual process using the rattling mode with higher energy at $\omega_{\ln} = 5$ meV.

Table IV. Key parameters characterizing the superconductivity of $\alpha$-pyrochlore $Cd_2Re_2O_7$ and three $\beta$-pyrochlore oxides $AOs_2O_6$.

|  | $\alpha$-$Cd_2Re_2O_7$ | $\beta$-$CsOs_2O_6$ | $\beta$-$RbOs_2O_6$ | $\beta$-$KOs_2O_6$ |
|---|---|---|---|---|
| $T_c / \omega_{\ln}$ [$\Delta C$] |  | 0.016 | 0.054 | 0.163 |
| $T_c / \omega_{\ln}$ [$H_c(0)$] |  | 0.046 | 0.103 | 0.168 |
| $\omega_{\ln}$ (K) [$H_c(0)$] | 458[a] | 70.7 | 61.0 | 57.1 |
| $\lambda$ | 1.63 | 2.76 | 3.38 | 6.3 |
| $\lambda^{sc}$ | 0.36 | 0.78 | 1.33 | 1.8 |
| $\lambda^{n}$ | 0.93 | 1.11 | 0.65 | 1.3 |
| $\lambda_{ep}$[b] |  | 0.77 | 0.80 | 0.85 |

[a]Debye temperature
[b]band structure calculations by Saniz and Freeman.[73]

*4.7 Electron-rattler interactions and mass enhancement*

We investigate the *e*-ph coupling constant $\lambda^{sc}$ responsible for the superconductivity using the well-known McMillan formula refined by Allen and Dynes:[90, 91]

$$T_c = \frac{\omega_{\ln}}{1.2} \exp\left[\frac{-1.04(1+\lambda^{sc})}{\lambda^{sc} - \mu^*(1+0.62\lambda^{sc})}\right], \quad (17)$$

where $\mu^*$ is the Coulomb coupling constant and is usually approximately 0.1. In fact, it has been estimated to be 0.096, 0.093, and 0.091 for the Cs, Rb, and K compounds, respectively.[73] Using the values of $\omega_{\ln}$ determined above, we estimate $\lambda^{sc}$ for Cs(Rb) to be 0.78(1.33). This value for Rb is significantly larger than the previously reported value of $\lambda^{sc} \sim 1$ by Brühwiler *et al.*[31] The difference originates from the different values of $\omega_{\ln}$; we used $\omega_{\ln} = 61.0$ K based on the strong-coupling correction of $H_c(0)$, while they used $\omega_{\ln} = 107$ K from $\Delta C$, which may be inappropriate owing to the gap anisotropy.

In our previous report on the K compound,[12] we estimated $\lambda^{sc}$ to be 2.38 using eq. (17) in the same manner. However, this may have overestimated the coupling constant, because eq. (17) is not rationalized for such an extremely strong coupling superconductor.[91] Alternatively, we estimate $\lambda^{sc}$ here on the basis of the strong-coupling correction of $\lambda^{sc}$. As shown in Fig. 25, $\lambda^{sc}$ is nearly proportional to $x$ as $\lambda^{sc} = 0.38(2) + 8.7(2)x$; thus, $\lambda^{sc} = 1.8$ is determined for K. The values of $\lambda^{sc}$ obtained for Cs and Rb using the McMillan-Allen-Dynes formula fall exactly on this line. However, for $Cd_2Re_2O_7$, one obtains $\lambda^{sc} = 0.36$ using the Debye temperature $\Theta_D = 458$ K instead of $\omega_{\ln}$ in eq.

(17). It is now clear that the large increase in $T_c$ by one order of magnitude from $Cd_2Re_2O_7$ to $KOs_2O_6$ is mostly due to the enhancement of $\lambda^{sc}$, which is large enough to compensate for the corresponding decrease in the phonon frequency. The origin of the enhancement of $\lambda^{sc}$ is attributed to the reduction of the phonon frequency because of the denominator of $\omega$ in eq. (13).

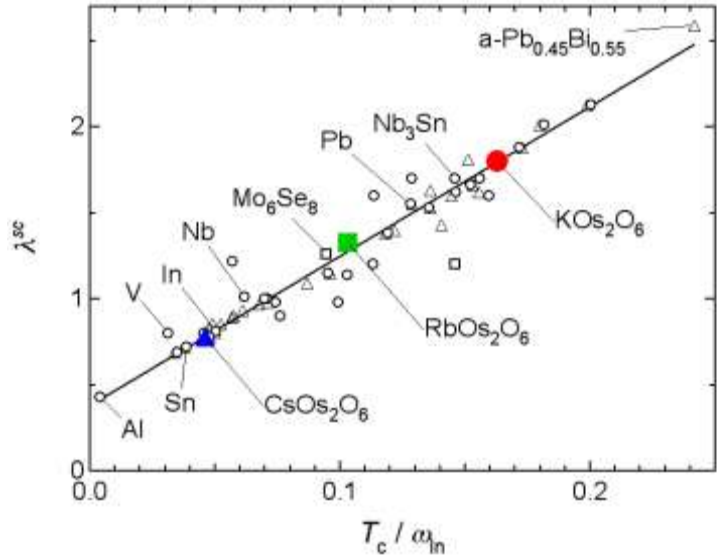

Fig. 25. (Color online) $e$-ph coupling constant $\lambda^{sc}$ vs $T_c / \omega_{ln}$. Materials shown by open circles and squares are the same as those in Fig. 24. Those shown by open triangles are from ref. 91. The straight line represents a fit to the linear relation of $\lambda^{sc} = 0.38(2) + 8.7(2)(T_c / \omega_{ln})$

Concerning the factor determining the $T_c$ of the $\beta$-pyrochlore oxides, we briefly mention the pressure dependence of $T_c$ thus far reported.[21, 92-95] For all the $\beta$-pyrochlore oxides, $T_c$ initially increases upon compressing the lattice, reaches a maximum, gradually decreases with further increasing pressure, and is finally suppressed completely above a critical pressure.[93] Although both the maximum value and the critical pressure depend on element $A$, a bell-shaped variation of $T_c$ is commonly observed. Saniz and Freeman found in their calculations that the effect of pressure on the overall electronic structures is rather small.[73] Thus, it is reasonable to assume that the coupling constant varies accordingly under pressure. The initial compression may enhance the coupling constant simply by decreasing the distance between the rattler and the electrons on the cage. On the other hand, further compression is thought to reduce the coupling constant, because the rattling frequency should increase with decreasing cage size; thus, the electron-rattler interactions are reduced. Thus, the pressure dependences of $T_c$ are interpreted qualitatively within the rattling scenario.

Next we discuss the mass enhancement in the normal state. The mass enhancement sometimes has a common origin such as the Cooper pairing in the superconducting state. As listed in Table IV and shown in Fig. 26, the mass enhancement factor $\gamma_{exp} / \gamma_{band} = 1 + \lambda$ increases enormously with increasing $T_c$. In particular, the large value of 7.3 for K cannot be explained by a single $e$-ph interaction, because $\lambda$ must be smaller than 2 in such a metallic system owing to the screening

effects.[96]) Thus, it is necessary to assume at least two independent renormalization processes; $1 + \lambda = (1 + \lambda^{sc})(1 + \lambda^{n})$, where $\lambda^{sc}$ affects the superconductivity and has been determined above, and $\lambda^{n}$, which results from the other effects. Interestingly, the thus-estimated $\lambda^{n}$ is always close to 1, as shown in Fig. 26. This means that there is already a mass enhancement by a factor of 2 common to the $\alpha$ and $\beta$-pyrochlore oxides, which is not responsible for the superconductivity. This additional enhancement may be attributed to an $e$-ph interaction on the cage or an $e$-$e$ interaction, or their combination. We have already mentioned, however, that there is no evidence for predominant $e$-$e$ scattering in the $\beta$-pyrochlore oxides; the observed $T^2$ behavior of $\rho$ for the $\beta$-pyrochlore oxides does not necessarily imply $e$-$e$ scattering. Moreover, the resistivity of $Cd_2Re_2O_7$ is proportional to $T^3$, not $T^2$, at low temperatures, suggesting that low-energy optical phonons are a dominant source of scattering.[2]) Thus, we consider that the origin of the commonly existing $\lambda^{n}$ is a normal $e$-ph interaction in the cage made of a transition-metal-oxygen framework. In fact, Saniz and Freeman calculated the $e$-ph coupling constant assuming a phonon with a Debye temperature of $\Theta_D = 285$ K from the cage and obtained $\lambda_{ep} \sim 0.8$ for all the $\beta$-pyrochlore oxides (Table IV),[73]) which is comparable to our value of $\lambda^{n}$. In this context, it is plausible to ascribe $\lambda^{sc}$ to the electron-rattler ($e$-r) interaction, that is, $\lambda^{sc} = \lambda_{er}$. The observed large enhancement of $\lambda^{sc} = \lambda_{er}$ toward K is clearly related to the corresponding decrease in $\omega_{ln}$, which is approximately equal to $\Theta_E$ as shown in Fig. 20. Thus, the electron-rattler interaction plays a crucial role governing all the electronic properties of the $\beta$-pyrochlore oxides.

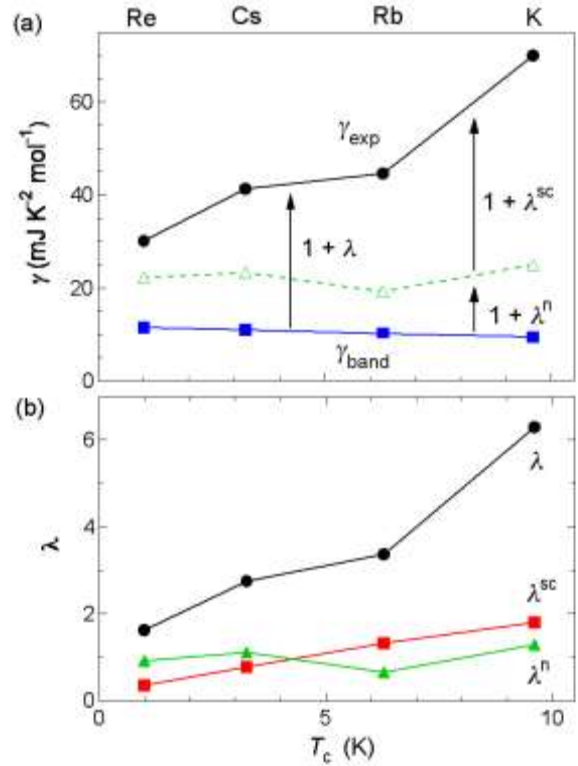

Fig. 26. (Color online) Evolution of the mass enhancement (a) and the coupling constants (b) over the series as a function of $T_c$.

A similar mechanism of mass enhancement due to rattling has been suggested for a filled skutterudite compound SmOs$_4$Sb$_{12}$, which exhibits a large value of $\gamma$ of 820 mJ K$^{-2}$ mol$^{-1}$, similar to that of other heavy fermion compounds.[97] Since $\gamma$ is insensitive to the magnetic field, it is considered that the origin is not due to an $f$ electron state, as in usual heavy fermion compounds, but to a possible interaction between conduction electrons and the rattling degree of freedom of the Sm atom. Hattori *et al.* studied an impurity four-level Kondo model, in which the Sm atom tunnels among four stable sites and interacts with surrounding conduction electrons, and found that the interaction can result in a moderate enhancement of the quasi-particle mass.[98] However, a related material, SmFe$_4$P$_{12}$, also shows a large, field-insensitive enhancement of $\gamma = 370$ mJ K$^{-2}$ mol$^{-1}$,[99] although the rattling is almost absent in this case owing to the small cage size compared with SmOs$_4$Sb$_{12}$. Moreover, $\gamma$ is much smaller in LaOs$_4$Sb$_{12}$, $\gamma = 36$ mJ K$^{-2}$ mol$^{-1}$,[100] in spite of the rattling of the La atom, similar to that of the Sm atom in SmOs$_4$Sb$_{12}$. Thus, it is not clear whether the large mass enhancement is due to the rattling degree of freedom, or how large the contribution of the rattling is. Further systematic study is necessary to elucidate the relation between the mass enhancement and the rattling in these skutterudite compounds.

*4.8 Related superconductors with low-energy phonons*

Finally, we discuss the characteristics of the $\beta$-pyrochlore superconductors in comparison with the other superconductors involving rattling or other types of low-energy phonons. In a recent extensive study on the filled skutterudite compounds, more than 10 superconductors were found in the family. PrOs$_4$Sb$_{12}$ with $T_c = 1.85$ K is the most attractive one, because its superconductivity seems to be unconventional.[101] In terms of the rattling, however, no clear relation has been established between $T_c$ or the nature of the superconducting state and the low-energy phonons in the skutterudite superconductors. For example, in the series of La$T_4X_{12}$ with $T =$ Ru, Os and $X =$ P, As, and Sb, $T_c (X)$ varies as 7.2 K (P), 10.3 K (As), and 3.58 K (Sb) for $T =$ Ru, and 1.8 K (P), 3.2 K (As), and 0.74 K (Sb) for $T =$ Os.[102] Accordingly, the Einstein temperature varies systematically as $\Theta_E(X) = 128$ K (P), 98 K (As), and 73 K (Sb) for $T =$ Ru and 131 K (P), 99 K (As), and 61 K (Sb) for $T =$ Os.[103] Although the rattling tends to become more intense with decreasing energy from $X =$ P to Sb in each series due to the larger cage, $T_c$ is always highest for As and lowest for Sb. Thus, such a simple relationship between $T_c$ and $\Theta_E$ as found in the $\beta$-pyrochlore oxides does not exist for the filled skutterudite compounds, implying that the superconductivity occurs by a different mechanism, or that other factors prevail.

Another class of compounds that exhibits rattling is the Si-Ge clathrates. Three superconductors have been reported thus far in the family: Na$_2$Ba$_6$Si$_{46}$,[104] Ba$_{24}$Ge$_{100}$,[105] and Ba$_{24}$Si$_{100}$[106] with $T_c = 4$, 0.24, and 1.4 K, respectively. Although one expects greater rattling for the germanate than for the silicates, $T_c$ is lower for the germanates. However, it is known that the Ba atom in Ba$_{24}$Ge$_{100}$ does not rattle at low temperatures, but locks randomly into split positions below $T_s \sim 200$ K. Moreover, $T_c$ markedly increases to 4 K as $T_s$ is suppressed by applying pressure.[105] This indicates that the resurgent rattling effectively enhances the $e$-ph coupling and thus increases $T_c$, though it was suggested in a previous work that the application of pressure reduces the strong disorder caused by the random displacement of the Ba atoms at $T_s$.[105] Nevertheless, similar to the skutterudites, the relation between the superconductivity and the rattling has not yet been established for the Si-Ge clathrates,

It is worth reconsidering "old" superconductors in the family of hexagonal tungsten bronzes $A_x$WO$_3$ with $A$ = Cs, Rb, K, and Tl with $T_c$ near 1.5 K.[107] They do not belong to the cage compounds but exhibit similar low-energy Einstein modes of $\Theta_E \sim 58$ K (Rb), 70 K (Cs), and 38 K (Tl), which originate from a rattling-like vibration in open channels along the $c$ axis. A recent INS study on the Cs and Rb compounds revealed the existence of local modes through the detection of a sharp peak at low energy in the phonon DOS.[108] However, no clear relationship between $T_c$ and the energy was observed, and, thus, it was suggested that the $A$ vibrations are not crucial for the superconductivity.

Another good example to be compared with the $\beta$-pyrochlore oxides is the family of Chevrel phase compounds with the ideal formula $A$Mo$_6$S$_8$, where, for instance, $A$ is Pb, Sn, Cu, Ag, and Yb with $T_c$ = 15, 14, 11, 9, and 8 K, respectively.[109, 110] They are regarded as "molecular crystals" with the $A$ atom weakly bound to channels made of the quasi-rigid Mo$_6$S$_8$ units. For the typical compounds of PbMo$_6$S$_8$ and SnMo$_6$S$_8$, low-lying Einstein modes at approximately 4.5 meV were found in INS experiments and ascribed to the optical mode of the Pb or Sn atom.[111] Interestingly, greater softening was found for the lighter Sn atom;[112] Sn is located above Pb in the periodic table. Moreover, a large-amplitude displacement of highly anharmonic character was found for the Sn motion by a Mössbauer study.[113] Thus, there are close similarities in terms of low-lying phonon modes between the Chevrel phase compounds and the $\beta$-pyrochlore oxides. It has been pointed out, however, that these low-lying modes have a minor influence on the $e$-ph coupling in the Chevrel phase compounds and that a soft phonon mode associated with the internal vibrations of the Mo$_6$S$_8$ unit is relevant to the mechanism of the superconductivity.[111, 114, 115] Nevertheless, it seems possible that the low-lying mode of the $A$ atoms couples with the soft mode of the Mo$_6$S$_8$ unit and thus enhances $T_c$ compared with binary compounds such as Mo$_6$Se$_8$ with $T_c$ = 6 K. In strong contrast, the present results for the $\beta$-pyrochlore oxides are rather straightforward, indicating clearly that the low-lying rattling modes induce the superconductivity through strong $e$-ph interactions.

It is well known that some classes of compounds exhibit strong-coupling superconductivity associated with various kinds of soft phonons. Typically, the well-known A-15 compounds exhibit a soft phonon mode associated with a cubic-to-tetragonal transition,[116] which can effectively enhance $T_c$. In addition, a marked variation of $T_c$ with the composition is found in Al$_{1-x}$Si$_x$ solid solution with the highest $T_c$ of approximately 11 K at $x = 0.2$.[117] This has been attributed to the enhancement of the $e$-ph interactions due to lattice instabilities in the nonequilibrium solid solutions. The role of such soft-mode phonons in the mechanism of superconductivity in these strong-coupling superconductors has been studied extensively.

## 5. Conclusion and Remarks

We have studied the thermodynamic properties of two $\beta$-pyrochlore oxide superconductors, $CsOs_2O_6$ and $RbOs_2O_6$, using high-quality single crystals. Reliable data on the superconducting and normal states have been obtained, which allows us to conclude that they are anisotropic $s$-wave superconductors. In addition, we discuss the chemical trends of various parameters over the series including $KOs_2O_6$ and $\alpha$-$Cd_2Re_2O_7$. The most important finding is that the mean phonon frequency derived from the strong-coupling correction analyses coincides with the energy of the $A$-ion rattling for each compound. This one-to-one correspondence is realized, because the rattling is essentially a local mode with low energy, giving a sharp peak in the phonon DOS. Therefore, we conclude that the rattling serves as an effective medium for the Cooper pairing in the $\beta$-pyrochlore oxides. The enhancement of $T_c$ and the characteristic change from weak- to strong-coupling superconductivity from Cs to K through Rb are attributed to the corresponding enlargement of the electron-rattler interactions with decreasing rattling frequency. By comparison to the related superconductors exhibiting rattling or other low-energy phonons, it is pointed out that the $\beta$-pyrochlore oxides present the first examples of compounds for which the rattling-induced superconductivity has been established.

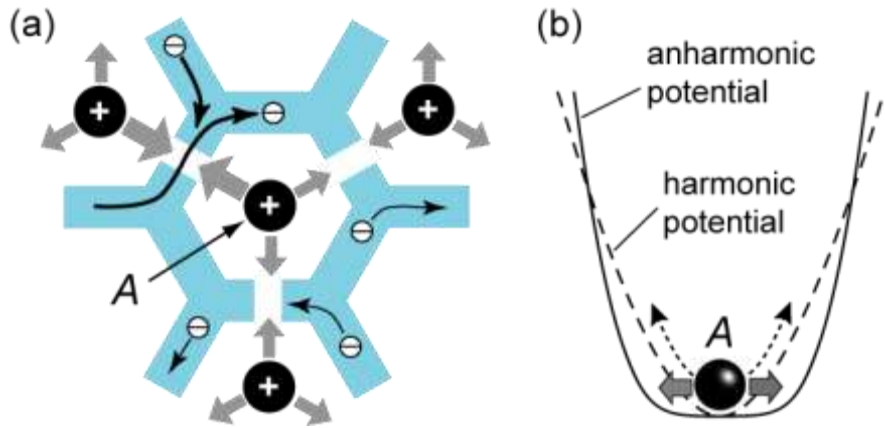

Fig. 27. (Color online) Schematic representation of the rattling-induced superconductivity in the $\beta$-pyrochlore oxides. When one electron is passing though the cage, nearby $A$ ions are attracted and move closer to each other, as depicted in (a) by thick arrows, with a large excursion from the center of the cage in such an anharmonic potential with a flat bottom as depicted in (b). Then, another electron is attracted by the excess positive charge induced by the movement of the $A$ ions before they return to their original positions. This is possible because the movement of the $A$ ions is much slower than that of the electrons and also because the restoring force is small owing to the anharmonicity. This process may cause an effective interaction between the two electrons in the real space. As the $A$ ions become smaller from Cs to K, the energy of the rattling decreases, and the anharmonicity increases. Accordingly, the pairing interaction increases, resulting in the distinctive variation from weak-coupling to extremely strong-coupling superconductivity from $CsOs_2O_6$ to $KOs_2O_6$.

Finally, we give a few remarks on the characteristics of the rattling-induced superconductivity in the $\beta$-pyrochlore oxides to be focused on in more detail in the future. One is its local nature. Distinct from the conventional weak-coupling phonon-mediated superconductivity, where a Cooper pair is assumed in the momentum space, it is realistic to assume paring in the real space for the $\beta$ pyrochlores, particularly for $KOs_2O_6$, because of the local nature of the rattling vibration as well as the small size of the Cooper pairs, as evidenced by the short coherence length. Thus, we can present the intuitive picture illustrated in Fig. 27. There may be substantial differences between the two types of superconductivity.

Another important issue is the role of anharmonicity. In most BCS-type superconductors, phonons have been tacitly assumed to be perfectly harmonic, although large-amplitude vibrations induced by a strong $e$-ph coupling should inevitably involve anharmonicity.[118] It is expected that strong-coupling superconductivity is influenced by anharmonicity. Yu and Anderson pointed out more than 20 years ago that large anharmonicity exists in the A-15 compounds and effectively enhances the $e$-ph interactions, giving rise to the increased $T_c$.[119] Nasu discussed the effects of anharmonicity by explicitly incorporating a local quartic potential and suggested that charge- or spin-density-wave instability is suppressed, stabilizing a superconducting ground state.[118] Curiously, a non-BCS-type isotope effect was predicted owing to the anharmonicity.[118] The $\beta$-pyrochlore oxides may be suitable compounds to study the effect of such anharmonicity on the mechanism of superconductivity, because the rattling with a large atomic excursion is essentially an anharmonic vibration. The isotope effect on $KOs_2O_6$ would be interesting to examine in future work.

A future INS experiment using a "large" single crystal would reveal the detail of the phonons in the $k$ space and the origin of the anisotropy in the superconducting gap or that of the electron-rattler interactions. Other experiments such as tunneling spectroscopy would be also helpful. In addition, first-principles calculations and theoretical treatments are required for understanding the nature of the electron-rattler interactions in more detail.

**Acknowledgments**

We wish to thank H. Harima, M. Ogata, T. Hasegawa, and H. Mutka for fruitful discussions. YN thanks H. Ueda for technical help in crystal growth. This research was supported by Grant-in-Aids for Scientific Research B (No. 16340101) and Scientific Research on Innovative Areas "Heavy Electrons" (No. 20102004) provided by the Ministry of Education, Culture, Sports, Science and Technology, Japan.